%% file: paper.tex
\documentclass[12pt, letterpaper]{JHEP3}
\usepackage{amssymb, amsmath, amsopn, amsthm}
\usepackage{epsfig}
\usepackage{psfrag}
\usepackage{subfigure}
\usepackage{cite}
\input texpref.tex

\newcommand{\mr}{\mathrm}

\def\be{\begin{equation}}
\def\ee{\end{equation}}
\def\bea{\begin{eqnarray}}
\def\eea{\end{eqnarray}}
\def\lp{\left(}
\def\rp{\right)}
\def\l{\left}
\def\r{\right}

\title{Holographic Gauge Mediation}

\author{Francesco Benini$^1$, Anatoly Dymarsky$^2$, Sebasti\'an Franco$^3$, Shamit Kachru$^{2,4}$, ~\\ ~\hspace{-.595cm}{\normalsize \bfseries \sffamily Dusan Simic$^{2,4}$ and Herman Verlinde$^{1,5}$}

~\\ ${}^1$Joseph Henry Laboratories, Princeton University,\\
Princeton, NJ 08544 USA \\ \vspace{0.3cm}

${}^2$Stanford Institute for Theoretical Physics, Stanford University,\\
Stanford, CA 94305 USA \\ \vspace{0.3cm}

${}^3$ Kavli Institute for Theoretical Physics, University of California,\\
Santa Barbara, CA 93106 USA \\ \vspace{0.3cm}

${}^4$ SLAC, Stanford University,
Stanford, CA 94309 USA \\ \vspace{0.3cm}

${}^5$Institute for Advanced Study,
Princeton, NJ 08540 USA \\ \vspace{0.3cm}

}

\newcommand{\bbea}{\begin{equation} \begin{aligned}} \newcommand{\eeea}{\end{aligned} \end{equation}}
\newcommand{\calS}{{\cal S}}

\abstract{
We discuss gravitational backgrounds where supersymmetry is broken at the end of a warped throat, and the SUSY-breaking is transmitted to the Standard Model via gauginos which live in (part of) the bulk of the throat geometry. We find that the leading effect arises from splittings of certain ``messenger mesons," which are
adjoint KK-modes of the D-branes supporting the Standard Model gauge group. This picture is a gravity dual of a strongly
coupled field theory where SUSY is broken in a hidden sector and transmitted to the Standard Model via a relative of semi-direct gauge mediation.
}

\preprint{PUPT-2292, SITP-09/01, SLAC-PUB-13500, NSF-KITP-09-18, ITEP-TH-05/09}

\def\be{\begin{equation}}
\def\ee{\end{equation}}
\def\bea{\begin{eqnarray}}
\def\eea{\end{eqnarray}}

\newcommand{\cF}{\mathcal{F}}

\newcommand{\cJ}{\mathcal{J}}

\newcommand{\cL}{\mathcal{L}}

\newcommand{\cN}{\mathcal{N}}

\newcommand{\cS}{\mathcal{S}}

\newcommand{\bZ}{\mathbb{Z}}

\newcommand{\spc}{\hspace{1pt}}
 
 \newcommand{\CC}{\mbox{\small $C$}}

\begin{document}


\tableofcontents

\section{Introduction}

The standard paradigm for supersymmetric model building invokes a hidden sector
where SUSY breaks, the MSSM (or some suitable extension), and a set of messenger
fields which transmit the SUSY-breaking to the MSSM.
For instance, in gravity mediation, the messengers are typically heavy fields which are integrated out at the string
or Planck scale, and induce cross-couplings between the field(s) carrying non-zero F-terms and
the observable sector.  In minimal gauge mediation, instead, the messengers are taken to be massive
particles $(\chi,\tilde{\chi})$ with Standard Model gauge charges, which feel SUSY-breaking through
coupling to a gauge singlet spurion $S$:
\bea
W_{mess} =  S \tilde\chi\chi, \qquad \quad S = M + \theta^2 F \, .
\eea
Gauge mediated scenarios have been a subject of great interest -- some classic references are
\cite{classic, classictwo} and an excellent review appears in \cite{GRreview}.    This interest
has largely been motivated by the fact that such scenarios
automatically incorporate a solution to the supersymmetric flavor problem -- the Standard Model
gauge fields couple to the different generations of quarks and leptons
in a flavor-blind way.  One recent direction
has been to consider generalizations of the minimal gauge mediation scenario, to include more
general possibilities for the messengers and their interaction with the hidden
sector.
Here, we consider the new possibilities that become calculable in light of gauge/gravity duality
\cite{GGD}.

Our interest is in supersymmetric phenomenological scenarios that, following the standard
paradigm, are specified by a Lagrangian of the factorized form
\bea
\label{threeterms}
{\cal L}  = {\cal L}_{\rm visible} + {\cal L}_{\rm hidden} + {\cal L}_{\rm int}\, .
\eea
We assume that the model fully fits within the framework of local 4D QFT, with a
UV cut-off at a high scale $\Lambda$ of order the GUT scale.
The three terms in ${\cal L}$ have the following minimal characteristics:\\[4mm]
${}$ \ \parbox{15cm}{\addtolength{\baselineskip}{.5mm}
$\bullet$ The visible fields constitute some supersymmetric extension of the Standard
Model. We assume that all (visible and hidden) matter particles are organized in
complete $SU(5)$ multiplets, so that coupling constant unification is preserved.
We denote the visible gauge multiplet by ${\cal V}$.}\\[4mm]
${}$ \ \parbox{15cm}{\addtolength{\baselineskip}{.5mm} $\bullet$ The hidden sector
is a strongly coupled supersymmetric gauge theory, that exhibits metastable SUSY
breaking at some scale $\Lambda_S >$ TeV. It has a global symmetry group $G$,
that contains $SU(5)$ as a subgroup. The corresponding current supermultiplet is
denoted by ${\cal J}(x, \theta, \bar \theta)$. }
 \\[4mm]
${}$ \ \parbox{15cm}{\addtolength{\baselineskip}{.6mm} $\bullet$
The interaction between the two sectors, at the
linearized level,  takes the form
\bea
\label{lint}
{\cal L} _{\rm int}  =  2 g_{\rm SM} \int \!\! d^{\spc 4}  \theta ~ 
 {\cal V} \spc {\cal J}
%
\eea
where $g_{\mr SM}$ denotes the SM gauge coupling.}\\

In this ``general gauge mediation"  formalism  \cite{meade-et-al}, the
soft masses are directly
extracted from 2-point functions of ${\cal J}$, via relations that do not depend on whether the
messenger fields actively participate in the strongly~coupled SUSY-breaking dynamics.

\smallskip

4D scenarios of the above general type admit a natural geometrization via the gauge/gravity correspondence.
The coupling (\ref{lint}) between the two sectors is of the right form to allow replacing the strongly
coupled hidden dynamics by the corresponding string dual.
Assuming that the hidden gauge group has sufficiently large rank and 't Hooft coupling, we may approximate
the dual theory as a classical supergravity theory living inside a warped 5D space time, cut-off at some radial
location $r = r_{\mr {UV } }$ (corresponding to the 4D cut-off scale $\Lambda_{\mr {UV } } $).
The presence of the global symmetry $G$  implies that the dual
5D supergravity theory contains a gauge superfield ${\cal V}$. Moreover, the
gauge/gravity dictionary identifies the classical supergravity action with fixed boundary values with the 4D
effective action obtained by integrating out the strongly coupled hidden sector gauge theory, with fixed values
of the SM gauge field ${\cal V}$:
\be
\label{GKPW}
\Bigl\langle e^{\mbox{\footnotesize $i \int  g {\cal V}{\cal J} $}} 
\Bigr\rangle_\mr{hidden} \, = \; e^{\mbox{\footnotesize $i S_{\rm sugra}({\cal V})$}}
\ee
%
The visible Lagrangian remains localized
at the boundary $r=r_{\mr {UV} }$, as an extra UV contribution to the total 4D action.  The SUSY-breaking dynamics
is localized in the IR region of the throat, and is communicated to the visible sector via the supergravity fields
that extend throughout the 5D bulk.\footnote{See for example \cite{Verlinde:2007qk,Dermisek:2007qi,Grimm:2008ed} for other string theory constructions in which a SUSY-breaking sector resides at the bottom of a warped throat and SUSY-breaking is mediated by a $U(1)$ gauge interaction.}

Normally in AdS/CFT, the asymptotic boundary values of the 5D fields act like non-dynamical sources.
However, with the cut-off in place, the internal space is compact, and the bulk fields have dynamical
boundary values and normalizable zero modes.
To extract the low energy physics, one could in principle use a combination of the
4D RG (for the boundary action) and a holographic RG (for the
classical bulk theory). A more direct route, however, is to apply the usual
Kaluza-Klein reduction.
In particular, the low energy MSSM gauge multiplet arises as the zero mode of the 5D vector superfield ${\cal V}$.

As in any phenomenological scenario that incorporates  hidden matter charged under the SM gauge
group, special care is needed to avoid a Landau pole. This issue
is particularly relevant in our construction, since the hidden sector is a large $N$ theory. The effective number of
extra charged matter fields shows up via the short-distance behavior of the 2-point function of the
scalar component of the supercurrent $\langle {J}(0) \, {J}(x)\rangle \sim \spc \CC/x^4$, which results in an
extra contribution $\Delta b_0 = - (2\pi)^4 \CC$ to the beta function of the MSSM gauge group.
From the 5D point of view, the extra contribution
to the RG running of  $1/g_{\mr {SM} }^2$ between two scales is just given by the integral of the 5D gauge field
action over the region between the corresponding radial positions. The Landau pole problem
thus  translates into an upper bound on the total radial range over which the 5D gauge field extends.

\medskip
\subsection{Basic setup and overview of results}

We now describe the specific string dual of 4D gauge mediation that we will study in detail in this paper.  It has the
following ingredients.

\smallskip

\noindent{\begin{itemize}
\item As a gravity dual, we take the warped deformed conifold geometry with a  small number $p$ of anti-D3 branes at its tip $r = \epsilon^{2/3}$.\footnote{For simplicity, we set $\alpha'=1$ in most of the paper. It is straightforward to re-introduce the necessary powers of $\alpha'$ when needed.} The warped conifold is the most explicitly known dual to a confining supersymmetric gauge theory \cite{KS}, and adding the anti-D3's amounts to placing the  theory in a metastable state,
that breaks SUSY at (due to the warping) an exponentially small energy $\Lambda_S$ \cite{Kachru:2002gs}.\footnote{There is a one-to-one correspondence between $\Lambda_{\cal S}$ and $\epsilon$. Because of this, we will often use them interchangeably. The same applies to $\mu$ and its associated scale $\Lambda_\mu$.}

\item We place $K$ D7-branes inside the
conifold geometry. In the gauge theory, this amounts to adding bifundamental messenger fields $(\chi,\tilde{\chi})$, that are charged under the hidden gauge group
and under the gauge group $G\! =\! SU(K)$ on the D7-branes, that contains the SM gauge~group. The radial location $r=\mu^{2/3}$ of the tip of the D7-branes sets the mass of the messengers, which we assume is
larger than the SUSY-breaking scale: $\mu^{2/3} > \epsilon^{2/3}$. The anti-D3 and D7 branes are thus geometrically separated.

\item We assume the chiral MSSM matter fields are localized at the UV end of
the geometry $ r = r_{UV}$.  In other words, they are ``elementary" matter fields, not composites formed by strong CFT dynamics.
As a result, their leading soft masses will arise by communicating with the bulk gauge multiplet, which lives in the throat (extending
down to $r =\mu^{2/3}$) and feels the leading effects of SUSY-breaking.

\end{itemize}}

\medskip

Note that the system is characterized by relatively few parameters: the SUSY-breaking scale and messenger mass, and some
discrete numbers such as the gauge group ranks. Besides setting up the model, our main task will be to try and use the supergravity to compute the soft masses of the MSSM particles.

Since the hidden and visible sector communicate via messenger fields $(\chi,\tilde{\chi})$, the gauge current ${\cal J}$ is
\be
\cJ =  \frac{\partial K }{\partial \chi_{a}} T_{ab}  \chi_b + \frac{ \partial K}{\partial {\tilde \chi_{a}} } T_{ab}  \tilde \chi_b + \dots
\ee
where $K(\chi, \tilde \chi, \bar \chi, \bar {\tilde \chi}, ..)$ is the K\"ahler potential of the theory and the $T_{ab}$ are the appropriate gauge matrices \cite{Weinberg_III}. The messengers
are in direct contact with, but do not take active part in, the
strongly coupled SUSY-breaking dynamics.
 From a 4D perspective, the model can thus be thought of as
a close relative of a ``semi-direct gauge mediation'' scenario \cite{semi-direct}\footnote{``Close relative"
because certain couplings in the holographic dual superpotential make the messenger
interactions with the hidden sector a bit
more direct, but seem to contribute only subleading effects in the gravity regime computations we
perform.}.
Below the hidden confinement transition, the messengers
are regrouped into ``messenger mesons'', of the general form\footnote{Besides scalar mesons, there are also vector messenger mesons.}
\be \label{messenger operators}
\Phi_n = \tilde \chi \spc {\cal O}_n\chi \, ,  \  \ \qquad
\ee
that transform in the adjoint representation under the MSSM gauge group. Here ${\cal O}_n$ denotes a hidden sector chiral operator, that transforms in the conjugate representation of both $\chi$ and $\tilde \chi$ with respect to the hidden gauge group.
We will see that it is the interaction of these messenger mesons with the MSSM gauge multiplet that
generates the leading soft masses in the observable sector.

The organization of this paper is as follows.  In \S2, we introduce the model in more detail.  In \S3, we compute the
leading SUSY-breaking mass of the messenger mesons in the metastable SUSY-breaking vacuum.   In \S4, we translate this
into a calculation of the gaugino mass and describe how the splittings in the gauge multiplet translate (via gaugino mediation
\cite{gaugino}) into soft masses for the MSSM matter multiplets; we also show that a more direct coupling of the gaugino to
the leading background supergravity solution which could cause a SUSY-breaking splitting is absent, so this semi-direct
mechanism is the leading effect.  In \S5, we discuss the issue of Landau poles.  We close
in \S6 with a description of future directions.  In appendix \ref{app: sugra solutions}, for completeness, we review some details of supergravity solutions that we
use in the bulk of the paper; in appendices \ref{DiracSection}-\ref{app: vector mesons deformed} we detail some of the lengthier calculations that
we refer to in the main body; in appendix \ref{app: slowing down} we consider some orbifolds of the conifold setup.

Scenarios closely related to the one we describe in the present work were previously considered, at a phenomenological level,
in \cite{Nomura}.  The goal of this work is in part to place these models in a more complete microscopic setting, and in part
to pave the way for future work involving additional dynamics, which changes both the conceptual picture and
the phenomenology in important ways \cite{toappear}.  This further work is motivated 
by the intriguing ideas about supersymmetric composite models in \cite{singlesector}.  Such models
are intrinsically strongly coupled, and gauge/gravity duality would offer one of very few techniques
to gain a quantitative handle on the strong dynamics.   Their 
gravity duals have been
explored, at a phenomenological level, in \cite{GGG}.  One can clearly interpolate between our
present model and such composite models by moving in some (or all) of the chiral matter from
$r = r_{\rm UV}$ to interior positions in the throat.  The present work thus provides the first step in
putting such dual descriptions of composite models on a firm foundation.

\bigskip
\section{The Model}

\begin{figure}[h]
\begin{center}
\psfrag{A12}[cc][][1]{$A_{1,2}$}
\psfrag{B12}[cc][][1]{$B_{1,2}$}
\includegraphics[width=5cm]{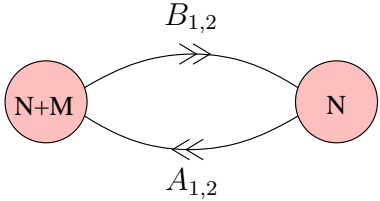}
\caption{Quiver diagram for the conifold.}
\label{quiver_conifold}
\end{center}
\end{figure}

In this section we describe in some more detail the model we focus on in the rest of the paper. As explained, we take the conifold theory as our large $N$ hidden sector. This theory has an $SU(N)\times SU(N+M)$ gauge group, which results from taking $N$ D3-branes and $M$ wrapped D5-branes (fractional branes) at the tip of the conifold. Its quiver diagram is shown in \fref{quiver_conifold}. The model has an $SU(2)\times SU(2)$ global symmetry, with the invariant superpotential given by

\beq
W=\epsilon^{ij} \epsilon^{kl}~ \Tr~ A_i B_k A_j B_l ~.
\label{W1}
\eeq
where the $i,j$ are $SU(2)$ indexes. 

The fractional branes break conformal invariance and the theory undergoes a cascade of Seiberg dualities, which gradually reduces the effective number of D3-branes and terminates in confinement at an exponentially small scale (compared to the UV scale) $\epsilon$ in the IR \cite{KS}. 
The gravity dual reflection of confinement is a solution which ends smoothly in
the IR, with a finite-sized
3-sphere characterizing the tip.

\begin{figure}[h]
\begin{center}
\psfrag{D3b}[cc][][1]{$\overline{{\rm D3}}$}
\psfrag{e}[cc][][1.3]{$\epsilon$}
\includegraphics[width=8cm]{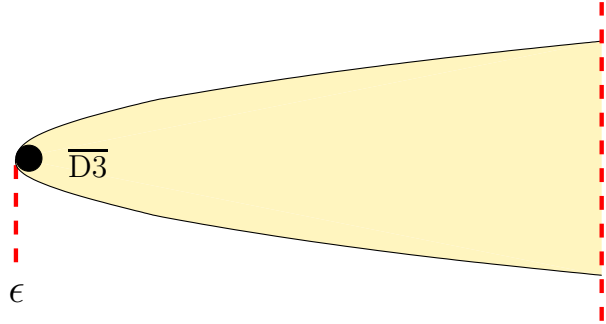}
\caption{Metastable $\overline{{\rm D3}}$-branes at the tip of the KS background.}
\label{D3bar}
\end{center}
\end{figure}

For $N=k M-p$ (with $k$ an integer) the hidden sector has a metastable, SUSY-breaking vacuum, with vacuum energy $\cS \sim p \, \Lambda_S^4$ \cite{Kachru:2002gs, DKM}. While a field theory understanding of the strongly coupled non-SUSY vacuum is still missing, it admits a clear description in the gravity dual. The $N$ D3-branes that are present in the UV gradually disappear and, after $k$ Seiberg dualities, we are left with $p$ anti-D3 branes. The stability of the anti-D3 branes and the tunneling decay to a SUSY vacuum with no anti-branes and $M-p$ D3-branes via annihilation against NS-flux has been studied in \cite{Kachru:2002gs, Lippert, Freivogel, Oliver}.  Of particular use to us will be the smeared supergravity
solution which captures the large $r$ physics of the metastable SUSY-breaking state \cite{DKM}.

The next step is to endow the hidden sector with a global symmetry. The SM gauge symmetry is a gauged subgroup of it. A global symmetry in 4d corresponds to a gauge symmetry in the bulk. In type IIB, we can introduce such bulk SM gauge fields via D7-branes that extend radially. The D3-D7 strings give rise to chiral fields in the (anti)fundamental representation of some of the conifold gauge groups. For this reason, such D7-branes are referred to as flavor branes.
For concreteness, we focus on realizing an $SU(5)$ gauge group. To do so, we introduce $K\ge 5$ flavor D7-branes. The model now has an $SU(K)$ global symmetry, the gauge symmetry on the worldvolume of the D7-branes, of which $SU(5)$ is a subgroup.

\begin{figure}[h]
\begin{center}
\psfrag{D3b}[cc][][1]{$\overline{{\rm D3}}$}
\psfrag{D7s}[cc][][1]{${\rm D7s}$}
\psfrag{e}[cc][][1.3]{$\epsilon$}
\psfrag{m}[cc][][1.3]{$\mu$}
\includegraphics[width=8cm]{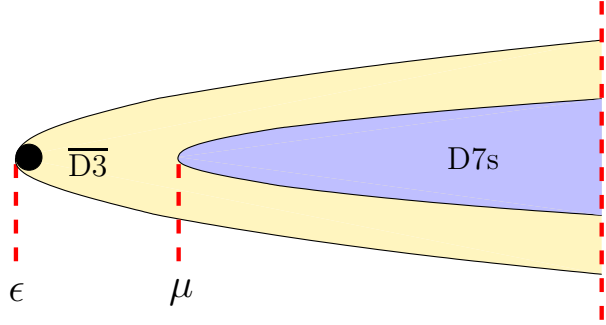}
\caption{The SM gauge fields live in the worldvolume of D7-branes.}
\label{D3bar_D7}
\end{center}
\end{figure}

\begin{figure}[h]
\begin{center}
\psfrag{A12}[cc][][1]{$A_{1,2}$}
\psfrag{B12}[cc][][1]{$B_{1,2}$}
\psfrag{chi}[cc][][1]{$\chi$}
\psfrag{chit}[cc][][1]{$\tilde{\chi}$}
\psfrag{in SU(5)}[cc][][1]{$\supset SU(5)$}
\includegraphics[width=8.5cm]{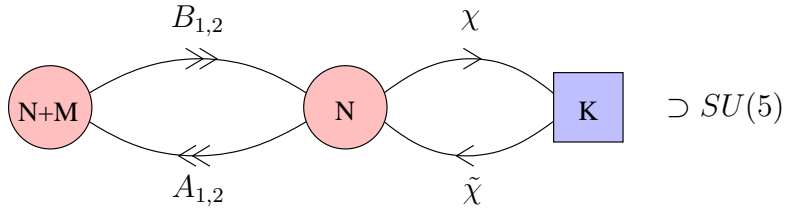}
\caption{Quiver diagram for the conifold flavored by Kuperstein D7-branes.}
\label{quiver_conifold_flavors}
\end{center}
\end{figure}

There are various ways in which D7-branes can be supersymmetrically embedded in the conifold,
which have been discussed in \cite{Ouyang,Kuperstein} and many subsequent papers.  For our purposes, realizing
the Standard Model gauge group on Kuperstein-embedded D7-branes \cite{Kuperstein} will suffice.  This embedding
is defined as follows. If we choose complex coordinates $z_i$ in which the defining equation of the deformed conifold geometry is
\be
\label{conifold}
\sum_{i=1}^4 z_i^2 = \epsilon^2,
\ee
then we embed a stack of $K$ D7-branes on the divisor defined by the equation
\be
\label{oureqn}
z_4 = \mu~.
\ee
The extended quiver which captures the field content of the gauge theory dual to $K$ such D7-branes in the
warped deformed conifold geometry is shown in \fref{quiver_conifold_flavors}.  In the non-compact throat, the $SU(K)$ gauge
group on the D7's is a global (flavor) symmetry group, and the additional matter fields are flavors in the $SU(N+M) \times SU(N)$ gauge
theory.  When the throat is glued into a compact Calabi-Yau manifold, the $SU(K)$ becomes weakly gauged.

The superpotential of the flavored theory becomes%
\footnote{This expression can be obtained from the $\cN =2$ parent theory, where the superpotential is fixed, upon mass deformation for the adjoint scalars. Strictly speaking, (\ref{W_flavors}) is correct for $U(N)$ groups while for $SU(N)$ there are $1/N$ suppressed double trace terms.}

\beq
\label{W_flavors}
W=\epsilon^{ij} \epsilon^{kl} A_i B_k A_j B_l + \tilde{\chi}^a \, (A_1 B_1+A_2 B_2) \, \chi_a + \mu \, \tilde{\chi}^a \, \chi_a +\tilde{\chi} \chi \tilde{\chi} \chi~,
\eeq
with $a=1,\ldots,K$. The $\chi_a$ and $\tilde{\chi}^a$ transform as bifundamentals of one of the hidden sector gauge groups and the
observable $SU(K)$ gauge group. We will refer to them as messenger fields.

We pause here to note that the messenger mass $\mu$ breaks R-symmetry, a necessary condition for non-vanishing gaugino masses.
Even in the absence of messengers, R-symmetry is broken in the IR at the scale $\epsilon$ by the dynamics associated with confinement.
$\mu$ can in principle be much larger than $\epsilon$. Then, the R-symmetry breaking due to \eref{W_flavors} is generally a higher scale effect. These statements have a simple geometric realization.
The equation defining the background geometry is (\ref{conifold}); the $U(1)$ R-symmetry (which is exact when $\epsilon \to 0$ but only
approximate for non-zero $\epsilon$) acts by rotating $z_i \to e^{-i \alpha} z_i$
From \eref{conifold}, we see that the R-symmetry is broken down to $\mathbb{Z}_2$ by the IR deformation. 
(It is also broken to $\mathbb{Z}_{2M}$ in the UV, by background flux values in the deformed
conifold solution \cite{OWitten}, but for $M>1$ this still forbids a gaugino mass).
More importantly,
any massive D7-brane embedding of the form \eref{oureqn} breaks R-symmetry completely. I.e., as
we said, R-symmetry is not only broken at low energies, but it is already broken at energies of order $\mu$.

In the non-compact geometry, the value of $\mu$ is a boundary condition. We also hold $\mu$ fixed (i.e. forbid massless
fluctuations of the adjoint scalars); in a fully detailed compact setting this should be accomplished through moduli stabilization
in the bulk of the Calabi-Yau manifold.
In the warped geometry, the lightest dynamical modes descending from the adjoint scalars will then be KK modes.  We now turn to
computing their
spectrum and their splittings in the metastable SUSY-breaking state.

In what follows, we adopt the gauge theory convention for the mass dimension of various parameters. We take $[\cS]=4$, in accordance with it being a vacuum energy. In addition, we take $[\epsilon]=[\mu]=3/2$. In \S{4.3}, we discuss these conventions in more detail.

\medskip

Before moving on, let us notice that if the operator $\Tr(A_1 B_1+A_2 B_2)$ develops an 
F-term in the SUSY-breaking state, the couplings in the superpotential (\ref{W_flavors}) lead to a tree-level non-SUSY splitting of the messengers (analogous to the one in minimal gauge mediation).  This non-$SO(4)$-invariant
F-term would preserve all symmetries of the metastable anti-D3 state, and 
would provide the leading contribution to the mass splitting among the messenger fields. If present, this would be the leading effect
which generates a gaugino mass term. However, we argue in \S4.2\ that any $SO(4)$ breaking
effects are suppressed (relative to what we compute) by powers of ${\epsilon^{2/3} \over \mu^{2/3}}$ .  Thus our scenario is similar to semi-direct gauge mediation. Moreover, this argument implies that 
 the smeared approximation of \cite{DKM} (which projects out all non-$SO(4)$ invariant terms), is sufficient to capture the dominant SUSY breaking physics for $\mu^{2/3}>\!\! >\epsilon^{2/3}$. 

\bigskip

\section{The Messenger Mesons}

In this section, we compute the supersymmetric masses and the SUSY-breaking soft masses of the tower of KK mesons, which are the
lightest states (other than the ${\cal N}=1$ gauge multiplet) living on the D7-branes.  Since the computations become somewhat
technical, we first describe the basic structure and goals of what we are doing, and reserve the fully detailed computations to appendices \ref{app: effective action} through \ref{app: vector mesons deformed}. 

As described above, we consider the Kuperstein brane embedding (with abelian gauge group, for now). We denote the KK multiplets corresponding to fluctuations of \eref{oureqn} by the chiral superfields $\Phi_n$ with $n$ indexing the tower. The leading interactions and soft terms of a mode in the tower are captured by 
\be
{\cal L} = \displaystyle{ \int d^4 \theta }~ \Phi_n \bar \Phi_n + \displaystyle{ \int d^2 \theta} ~X_n \Phi_n \Phi_n + c.c.  ~,
\label{efft}
\ee
with $X_n = M_n + \theta \theta~F_n$. From \eref{oureqn} it is easy to see that the $\Phi_n$ carry positive unit R-charge\footnote{The R-symmetry is geometrically realized as a rotation of the conifold coordinates $z_i \rightarrow e^{i\theta} z_i$.}, and hence any R-breaking is entirely due to nonzero $F_n$. Loops of KK modes can generate a non-zero gaugino mass once $F_n \neq 0$\footnote{SUSY-breaking but R-neutral terms cannot contribute to the gaugino mass at leading order, and this is why we have omitted terms like $\l [ X \bar X \Phi \bar \Phi \r ]_D$ in \eref{efft}.}. 

The exact computation of the messenger mesons' action in the warped-deformed conifold is rather complicated (we compute their spectrum in the appendices).  However, for $\mu \gg \epsilon$, it
suffices to compute them in an approximate SUGRA solution which captures the large $r$ behavior of the geometry.  In the
supersymmetric vacuum state, such a solution was constructed by Klebanov and Tseytlin \cite{KT}; we henceforth refer to it
as the ``KT" solution.  The asymptotic supergravity solution corresponding to the metastable SUSY-breaking states was
determined in \cite{DKM}; we will refer to this as the ``DKM" solution.  We will warm up by computing the spectrum of
meson masses $\{M_n \}$ in the KT solution, and then proceed to find the $F_n$ by studying probe D7-branes in the DKM solution.  Relevant details about these supergravity solutions are summarized in appendix A.

The full computations even in the asymptotic backgrounds are still rather involved. Instead of discussing the full computation in the main text, we present in \S3.1 a simplified calculation which gives the correct basic
physics for the supersymmetric spectrum of mesons; in \S3.2 we include the DKM correction to the asymptotics (due to SUSY-breaking) and
find the mesonic soft masses; and in \S3.3 we discuss some important qualitative features of the full KK tower gleaned from the more detailed computations presented in the appendices.

\medskip

\subsection{Supersymmetric meson masses}

Let us, to begin with, consider probe D7-branes in the KT background (whose details are presented in appendix \ref{app: sugra solutions}).
This will allow us to check our simplified calculation
via comparison to \cite{Kuperstein, LO}. The simplified approach is to  consider the following truncated expansion of the D7-brane action:

\beq
\label{original}
S = -\frac{\mu_7}{g_s}\int d^8 \sigma \sqrt{|\gamma|}~ \lp g_{4\bar 4} ~ \gamma^{ab}  \partial_a X^4 \partial_b X^{\bar 4} + \gamma^{ac} \gamma^{bd} F_{cb} F_{da} \rp \, ,
\eeq
where $X^4$ corresponds to small fluctuations of \eref{oureqn}. The metric $\gamma_{ab}$ is the induced metric on the brane and $g_{4 \bar 4}$ is the $4\bar 4$ component of the KT metric in the $z^i$ coordinate system. Note the absence of $g_{44}$ and $g_{\bar 4 \bar 4}$ terms.  They vanish in KT, but are nonzero in DKM, and are ultimately what lead to a gaugino mass. We discuss this in the next section. 

Denote the Minkowski coordinates by $x$ and the internal ones by $y$. We expand $X^4$ in KK modes

\beq
X^4 = \sum_{n} \phi_n(x)  \, \xi_n(y) \, ,
\eeq
where the $\xi_n$ satisfy 
\beq
\partial_b \lp \sqrt{|\gamma|}~  g_{4\bar 4} ~ g^{ab}  \partial_a \xi_n(y) \rp = - \lambda^2_n \xi_n(y) \, .
\label{eom}
\eeq
As discussed previously, we choose our boundary conditions at the end of the throat (where we imagine the compactification ``starting") so as to mimic a D7 wrapping a rigid 4-cycle. Operationally this means that the spectrum is gapped, $\lambda^2_n > 0$. Since the KT background has 4d ${\cal N} =1$ SUSY, we obtain upon dimensionally reducing \eref{original} the action in \eref{efft} with $F_n = 0$. In order to estimate $M_n$, it suffices to notice that \eref{original} implies 

\be
M^2_n = \frac{ \displaystyle \int d^4y\, \sqrt{|\gamma|} ~  g_{4 \bar 4} ~  \gamma^{ab} ~ \partial_a  \xi ~\partial_b \bar \xi}{ \displaystyle \int d^4y\,  \sqrt{|\gamma| }  ~h(r) ~ \tilde g_{4 \bar 4} ~ \xi \bar \xi }
\ee
where $h(r)$ is the KT warp factor and tildes signify the un-warped metric. In KT we have
\beq
h(r) = \frac{L^4_{\mr{eff}}(r)}{r^4} , \quad \tilde g_{4 \bar 4 } \sim \frac{1}{r}~
\eeq
where $L_{\mr{eff}}^4(r) \sim  4 \pi g_s N \alpha'^2 \ln(r/r_s)$. In terms of the hidden sector effective 't Hooft coupling $\lambda_\mr{eff}(r)=g_s N_\mr{eff}(r)$, we have $L_{\mr{eff}}^4(r)= 4 \pi \alpha'^2 \lambda_\mr{eff}(r)$. Now we make the replacement
\beq
\int d^4 y \sqrt{|\gamma|} \;\rightarrow\; \int_{|\mu|^{2/3}}^{r_{_{\rm UV}}} dr ~r^3~.
\eeq
Assuming the $\xi_n$ are well-localized around $r = |\mu|^{2/3}$ and by switching the integration variable to $x = r/|\mu|^{2/3}$ we find that\footnote{When deriving \eref{mesopprox} and similar expressions that follow, we have re-introduced the missing powers of $\alpha'$ appropriately.}
\beq \label{mesopprox}
M_n \propto \frac{|\mu|^{2/3} }{\sqrt{4 \pi \lambda_\mr{eff}(\mu)}} \, .
\eeq
Notice that the mesons are much lighter then their constituent messenger fields $\chi$, $\tilde\chi$. The messengers have mass $m_\chi = \mu^{2/3}$, while the lightest meson mass and the mass spacing in the tower are smaller by a factor of $(g_s N_\mr{eff})^{-1/2}$. Apparently, the strong dynamics of the conifold theory provides a large binding energy. The above qualitative estimates are confirmed via a more detailed calculation, presented in appendix E.

\subsection{Inclusion of SUSY-breaking}

The DKM solution provides a SUSY-breaking correction to \eref{original}:
\be
\label{orig_DKM}
\delta S_\mr{DKM} = -\frac{\mu_7}{g_s}\int d^8 \sigma \sqrt{|\gamma|}~ \Big(  g_{44} ~ g^{ab}  \partial_a X^4 \partial_b X^{ 4} 
+ g_{\bar 4 \bar 4} ~ g^{ab}  \partial_a X^{\bar 4} \partial_b X^{ \bar 4}  \Big) \, .
\nonumber 
\ee
The functions $g_{44} = (g_{\bar 4 \bar 4})^*$ are given by
\be
\tilde g_{44} = h^{-1/2} g_{44} \sim \bar \mu^2 \frac{ \cS}{r^8} \, ,
\label{DKMcorrection}
\ee
where ${\cal S}$ is the SUSY-breaking order parameter (the vacuum energy sourced by the anti-D3 branes) in the DKM solution\footnote{The rapid
fall-off of the perturbation visible in (\ref{DKMcorrection}) is consistent with the fact that these are ${\it normalizable}$ perturbations
to the conifold geometry, characterizing a ${\it state}$ in the supersymmetric field theory.} and as usual tilde denote un-warped quantities. The non-zero $g_{44}$ will source the $F_n$ in the 4d effective theory. By thinking about the KK reduction of \eref{orig_DKM} to 4d we discern

\beq
F_n = \frac{ \displaystyle \int d^4y\, \sqrt{|\gamma|} ~  g_{4  4} ~  \gamma^{ab} ~ \partial_a  \xi ~\partial_b  \xi}{ \displaystyle \int d^4y\, \sqrt{|\gamma| }  ~h ~ \tilde g_{4 \bar 4} ~ \xi ~ \bar \xi }~.
\label{u^2}
\eeq
Since the subsequent analysis is completely parallel to the analysis for $M_n$, we will just state the result (details appear in appendix D):
\beq
F_n \propto \frac{\bar \mu^2 \cS}{|\mu|^{10/3} 4 \pi \lambda_\mr{eff}(\mu)} \, .
\label{F_n}
\eeq
Note that this expression for $F_n$ manifestly displays the correct R-charge of minus two, and that the power-law fall off of $F_n$ as $\mu \rightarrow \infty$ is consistent with the restoration of SUSY in the UV. 

\subsection{More detailed computation of the spectrum} 

The above heuristic analysis gives the correct parametric dependence of the meson masses and soft terms on the parameters of the solution. However, it says nothing about \emph{how many} mesons there are localized at $r = \mu^{2/3}$, nor how their energy levels and SUSY-splittings are spaced. This will be important when summing the meson tower contribution to the gaugino mass. A more complete numerical analysis, summarized in appendix D, indicates that 
\be
M_n \propto n \frac{|\mu|^{2/3} }{4 \pi \sqrt{\lambda_\mr{eff}(\mu)}}  , \ \  F_n \propto n^2 \frac{\bar \mu^2 \cS}{|\mu|^{10/3} 4 \pi \lambda_\mr{eff}(\mu)}~.
\label{exact}
\ee

Naively one might have expected $F_n$ to \emph{decrease} with $n$, due to the fact that in a system with spontaneously broken SUSY, one expects SUSY restoration at high energies. However, probing greater \emph{total} energy is not the same as probing \emph{shorter} length scales. A state localized at the holographic radius $r$ probes a length scale in the gauge theory given by  \cite{Sussk_Witten, Polch_Peet}:
\be
l  = \frac{ \sqrt{g_sN_{ \mr {eff} } }  }{r}
\ee
even though its total energy may well exceed $1/ l$. This is precisely applicable to the mesons, as all relevant states are located at the same holographic scale $r = \mu^{2/3}$ and therefore probe the same length scale in the dual gauge theory.  In fact, using scale/radius duality, one can see that the $n$th
excited meson corresponds to a state whose mass is $\sim n$ times the cut-off scale, were we to
cut the theory off at $r = \mu^{2/3}$ \cite{Polch_Peet}.  If we think of the meson as composed of $n$ elementary 
quanta, each with a fixed splitting $\sqrt{F}$, then it should not be surprising that its
own splitting grows linearly with $n$.  This is precisely what we have found.

 \medskip

\noindent
{\it Remarks}

\begin{itemize}
\item {It's worth emphasizing that there are entire towers of meson localized at radii $r  > \mu^{2/3}$, which we are ignoring. For the purpose of computing SM soft terms, this is justified, since the mesons localized at a scale $r = \nu^{2/3} >\!\! > \mu^{2/3}$ will feel an F-term of order $ \frac{ \cS}{|\nu|^{4/3} \lambda_\mr{eff}(\nu)} <\!\! < \frac{\cS}{|\mu|^{4/3} \lambda_\mr{eff}(\mu)}$. 
}
\end{itemize}
 
\bigskip

\newcommand{\re}{\,\mathbb{R}\mbox{e}\,}
\newcommand{\im}{\,\mathbb{I}\mbox{m}\,}
\newcommand{\hyph}[1]{$#1$\nobreakdash-\hspace{0pt}}
\newcommand{\dvol}{d\text{vol}}
\providecommand{\abs}[1]{\lvert#1\rvert}
\newcommand{\Nugual}[1]{$\mathcal{N}= #1 $}
\newcommand{\sub}[2]{#1_\text{#2}}
\newcommand{\ib}{\bar \imath}
\newcommand{\jb}{\bar \jmath}
\newcommand{\doubt}[1]{\noindent $\clubsuit$ #1 $\clubsuit$}
\newcommand{\parfrac}[2]{\frac{\partial #1}{\partial #2}}
\newcommand{\orders}[1]{\calO \Bigl( #1 \Bigr)}
\newcommand{\ntitle}[1]{\begin{center} \LARGE \textbf{#1} \end{center}}

\numberwithin{equation}{section}

\newcommand{\nn}{\nonumber}
\newcommand{\tr}{\mbox{tr}}    
\renewcommand{\bea}{\begin{equation} \begin{aligned}}
\renewcommand{\eea}{\end{aligned} \end{equation}}

\newcommand{\calF}{\mathcal{F}}
\newcommand{\calO}{\mathcal{O}}
\newcommand{\calM}{\mathcal{M}}
\newcommand{\calV}{\mathcal{V}}
\newcommand{\calC}{\mathcal{C}}
\newcommand{\bbZ}{\mathbb{Z}}
\newcommand{\bbC}{\mathbb{C}}
\newcommand{\bbP}{\mathbb{P}}
\newcommand{\bbN}{\mathbb{N}}
\newcommand{\bbR}{\mathbb{R}}
\newcommand{\spartial}{{\bf \slash} \!\!\! \partial}
\newcommand{\sD}{\,{\bf \slash} \!\!\!\! D}

\section{Soft Terms}

In this section, we estimate the SUSY-breaking soft terms in our scenario.
Such terms are computed by integrating out the physics of the hidden sector. We can split the computation into two parts, by integrating above and below the energy scale $\Lambda_\mu = \mu^{2/3}$ set by $\mu$. Above $\Lambda_\mu$ the hidden plus messenger sector is a complicated strongly coupled theory, whose contribution to the gaugino mass is computed holographically by supergravity. The gaugino is a fermionic KK mode of the D7 Lagrangian, whose wavefunction was identified in the literature \cite{CIU, MMS}, and its effective mass at scale $\Lambda_\mu$ is induced by the coupling of the worldvolume action with the background.

Below $\Lambda_\mu$, the effective theory the visible sector is coupled to is a weakly coupled theory of mesons (almost free in the large $N$ limit \cite{'tHooft:1973jz}). The fact that the D7-branes end at $r=\mu^{2/3}$ tells us that the visible sector is essentially decoupled from the KS dynamics below $\Lambda_\mu$, and its only remnant is the tower of mesons. Since they are weakly coupled, we can compute their contribution to the gaugino mass at 1-loop.

In \S4.1, we prove that
no tree-level gaugino mass is induced by plugging the D7 probe action into the DKM solution (though
many of the calculations are relegated to appendix B).  In \S4.2, we provide macroscopic reasoning
to estimate the a priori expected gaugino mass (and justify the use of the smeared solution
\cite{DKM} in the computation). 
We proceed to directly compute the leading contribution to the gaugino mass in \S4.3; it is generated by
messenger loops, and is in accord with our estimate of \S4.2.  In \S4.4, we then determine the matter soft masses arising from gaugino mediation \cite{gaugino}:
the gaugino propagates up the throat geometry and
transmits its SUSY-breaking mass to the UV-localized matter fields.

\subsection{Absence of gaugino mass in supergravity}
\label{S4.1}
Before calculating the tree-level gaugino mass in the DKM case let us briefly review the
results applicable for the supersymmetric torus compactification with imaginary self-dual
(ISD) flux \cite{CIU}. The ten-dimensional Majorana-Weyl spinor
$\theta$ can be decomposed into the representations of the holonomy $SU(3)$ group ${\bf 0}+{\bf 3}$.
At each point the singlet and two of three vector states are massless and the third vector state is massive.
Clearly in the conifold case the definition of massive and massless vector modes will depend on location, but the singlet is well-defined globally. In the SUSY case the singlet state is massless and we identify it with gaugino $\lambda$.

On a more technical level, as we described in appendix B, the mass of  $\lambda$
is sourced by the $(0,3)$ flux
\bea
\label{z3}
Tr(\lambda^2)(G^3_{{\bar 1}{\bar 2}{\bar 3}})^*\ ,
\eea
and by the $(3,0)$ flux $Tr(\lambda^2)G^3_{ 123}$ if a non-trivial D7-brane world-volume flux is present.
There is also a coupling between $\lambda$ and the vector states $\Psi_I$. Thus 
in the case of torus compactification \cite{CIU} there is coupling to the vanishing 
imaginary anti-self dual (IASD) $(1,2)$ flux
\bea
Tr(\lambda \Psi^a)  (A_{\bar a})^* \ ,
A_{ a}= G^3_{a  b \bar c} \, g^{b \bar c}\, .
\eea
In the conifold case there might be other couplings but they all must vanish if SUSY is not broken. 

To find the gaugino mass in the SUSY-breaking DKM background one would need to compute the mass matrix for all four components of $\theta$ and diagonalize it. Luckily, we are interested in the leading correction to the unperturbed result.
Therefore we can use the definition of the unperturbed gaugino wave function as a singlet mode and compute the coefficient in front of $Tr{\lambda^2}$ at the leading order in ${\cal S}$. To this end we need to consider how (\ref{z3}) would change under the perturbation of the 3-form flux, background and induced metrics.

As we are working at the leading order in ${\cal S}$ we can consider the perturbations of flux and metric independently. Clearly the perturbation of the 3-form will not contribute as the DKM solution does not have $(0,3)$ or $(3,0)$ flux. Similarly the $(1,1)$ perturbation of the metric cannot change the flux type and hence does not contribute. The only non-trivial contribution may come from the $(0,2)$ perturbation of the 
bulk metric $\delta g^{{\bar a}{\bar b}}$ or the combination of the induced metric and the world-volume flux $\delta (M^{-1})^{{\bar \alpha}{\bar \beta}}$.
The latter sources a coupling of the gaugino to the ISD $(1,2)$ flux
\bea
 \epsilon_{\bar{\alpha}\bar{a}\bar{b}}\delta (M^{-1})^{({\bar \alpha}{\bar \beta}) }g^{{\bar a}c}g^{{\bar b}d}(G^3_{ \beta { \bar c}{\bar d}})^*\ ,
\eea
and ISD $(1,2)$ and IASD $(2,1)$ fluxes
\bea
 \epsilon_{\bar{\alpha}\bar{a}\bar{b}}\delta (M^{-1})^{[{\bar \alpha}{\bar \beta}] }g^{{\bar a}c}g^{{\bar b}d}\left((G^3_{ \beta { \bar c}{\bar d}})^*-G^3_{\bar \beta c d}\right)\ .
\eea
Recall that the fluxes $G$ appearing here are the ${\it unperturbed}$ ones.
Clearly these terms do not contribute in a near-KS backgrounds like DKM, since the KS background has
purely ISD (2,1) flux.  In fact, no Calabi-Yau background (with sufficiently generic holonomy) has 
non-trivial IASD (2,1) forms or ISD (1,2) forms.

The perturbation $\delta g^{{\bar \alpha}{\bar \beta}}$ sources two terms:
\bea
\label{M1}
 \epsilon_{\bar{a}\bar{b}\bar{c}}\delta g^{\bar{a}\bar{d}}g^{\bar{b}e}g^{\bar{c}f} G^3_{\bar{d}ef}\ ,
\eea
and
\bea
\label{M2}
 \epsilon_{\bar{\alpha}\bar{a}\bar{b}}\gamma^{\bar{\alpha}\beta}\delta g^{\bar{a}\bar{c}}g^{bd} \left((G^3_{\bar{\beta} c \bar{d}})^*-G^3_{\beta \bar{c}d}\right)\  .
\eea
Both are a priori potentially non-zero. Nevertheless, after substituting the DKM metric and the 3-form flux
into (\ref{M1}) and (\ref{M2}), they both vanish (see appendix \ref{DiracSection} for details).

\subsection{Expected parametric dependence of gaugino mass}

Before computing the result, it is useful to summarize our expectations for the gaugino mass.
How will it depend on the standard model coupling $g_{\mr{SM}}$, 
the SUSY-breaking order parameter ${\cal S}$, and the $\mu$ parameter of the D7-embedding?

\medskip
\noindent
$\bullet$  We expect the leading gaugino mass to be
proportional to $t = g_{\mr {SM} }^2 K$.   This is because the hidden sector communicates to the D7 gauge
theory only via the messengers $\chi$ and $\tilde \chi$, so the SUSY-breaking must vanish as 
$g_{\mr{SM}} \to 0$, and the messengers come in the $K$ dimensional representations of the D7 gauge group. More precisely, we expect this to be a one-loop contribution implying an additional factor of $\frac{1}{16 \pi^2}$.

\medskip
\noindent
$\bullet$ To argue for the correct ${\cal S}$ dependence, we use a strategy from \cite{MN}.
The anti-D3 source perturbs the supergravity equations in two ways: by sourcing a 
tension-term in the Einstein equations, and perturbing the RR gauge field $C_4$.  However,
we can imagine starting with a supergravity solution with an ``imaginary brane" source at
the location of the anti-D3.  The ``imaginary brane" has negative D3 charge and negative
D3 tension; it therefore preserves the same SUSY as the background, and generates no
gaugino mass.  The anti-D3 state differs from the ``imaginary brane" background by
sourcing two additional units of (warped) D3 tension, but the coupling to the $C_4$ field
remains unchanged.  Hence the only perturbation to the supergravity equations, on top
of a theory where the gaugino has vanishing mass, involves the parameter ${\cal S}$ to
the first power.  We therefore expect that the gaugino mass will be proportional to ${\cal S}$.

\medskip
\noindent
$\bullet$  R-charge considerations then fix the power of $\mu$; there must be a power of
$1/\mu^2$ to compensate the R-charge of $\cal S$.

\medskip
\noindent
Putting these considerations together, we expect a gaugino mass that scales like
\begin{equation}
\label{expect}
m_{\lambda} \sim {t\over 16\pi^2} ~ {{\cal S} \over \mu^2}~.
\end{equation}
Because at strong 't Hooft coupling for the hidden sector one is not always sure of the
expected powers of $\lambda_{\mr{eff}} = g_sN_{\mr{eff}}$, we will remain agnostic about that until
performing the direct calculation.

How large should the corrections to (\ref{expect}) be?  The leading corrections are 
related to the fact that the DKM solution is a smeared solution.  In a microscopic solution
with properly localized anti-D3 SUSY-breaking, one would expect a power series of
corrections to (\ref{expect}) that are R-neutral and give some function $F(\epsilon/\mu)$
that vanishes as $\epsilon \to 0$.  The fact that such corrections exist is evident already
from the computations in \S3 of \cite{DKV}, where some of the perturbations to the
supergravity due to localized anti-D3 sources are computed in the near-tip region.
These corrections are due to a dipole effect: the anti-D3 branes polarize the background five-form flux,
and as a result, source supergravity fields that they do not directly source via their
DBI action.  However, such effects (as expected for dipole effects) are both proportional
to the size of the dipole (and hence $\epsilon$) and fall off rapidly as one moves away
from the tip.

Finally, we should note that in the dual field theory, one would expect from 
(\ref{W_flavors}) that there would be a one-loop correction to the D7 gauge coupling function
$\sim \Tr AB$, giving rise to a potential gaugino mass from $F_{AB}$.  At least heuristically, the
computations in \cite{DKV} appear to show that in the gravity regime, these effects are
suppressed by powers of ${\epsilon\over \mu}$ compared to the effect we calculate.

\subsection{Loopy gaugino mass}
\label{S4.3}

In \S3.2, we derived an action for the messenger mesons, $\Phi_n$,
including the SUSY-breaking effects of DKM. Gauge invariance and SUSY allow
us to non-abelianize \eref{efft}
\be
{\cal L} = \displaystyle{ \int d^4 \theta }~ \Tr \{ \Phi_n e^{-V} \bar \Phi_n \} + \displaystyle{ \int d^2 \theta} ~X_n ~\Tr \{\Phi_n \Phi_n \} + c.c.  ~, 
\label{eff}
\ee
where now the $\Phi_n$ are adjoints of $SU(N_c)$. The one-loop contribution to the gaugino mass by the $n$-th meson is given by the standard gauge mediation formula (with the $\Phi_n$ performing the role of the messengers)
\beq
\label{gmass}
 \frac{t}{16\pi^2} \,  \frac{F_n}{M_n} \;\sim\; \frac{t}{16\pi^2} \,  \frac{n~\cS}{\mu^2 \sqrt{4 \pi\lambda_\mr{eff}}}
\eeq
where, again, $t$ is the 't Hooft coupling of the MSSM gauge theory on the D7-branes.
 The actual gaugino mass is obtained by summing the effect of all mesons:
\be
m_\lambda = \frac{t}{16\pi^2} \,  \frac{\cS}{\mu^2 \sqrt{4 \pi \lambda_\mr{eff}}} \sum_{n} n ~e^{i\theta_n}\;.
\ee
We have made the phases in the sum explicit to emphasize that there could be phase cancellations due to the fact that the $F_n$ are generally complex.  Notice that up to the numerical factor coming from
the sum over the tower, and powers of the hidden sector 't Hooft coupling, this is in accord with
our expectations (\ref{expect}).

\smallskip 

Although naively this sum looks divergent, it is important to realize that the description in terms of mesons eventually terminates. A strict upper bound is placed on the sum by the binding energy of the mesons, which is $\Lambda_\mu = \mu^{2/3}$, corresponding to the bound  $n <  (g_s N_\mr{eff})^{1/2}$. Above this scale the mesons de-confine and the full QFT dynamics including the messenger fields $(\chi, \tilde \chi)$ re-couples. However, an even lower upper bound might arise due to the fact that around $n = (g_s N_\mr{eff})^{1/4}$ the DBI expansion breaks down. In fact this is around when the mesons have masses of order those of massive closed string states localized at $r=\mu^{2/3}$. Thus we have
\be
(g_s N_\mr{eff})^{1/4} < n_\mr{max} < (g_s N_\mr{eff})^{1/2}.
\ee
More thought is required to determine $n_\mr{max}$ more precisely, but if we work in the regime of greatest interest with moderate values of the control parameters
(e.g. $4\pi g_s N \sim 10^2$) this rough understanding will suffice for our estimates of soft masses.

\bigskip
\noindent
{\it Remarks}
\begin{itemize}
\item {It is worth explaining the conventions in the formula (\ref{gmass}). As we have already mentioned, one should interpret $\mu^2$ as having
dimensions of mass$^3$ and ${\cal S}$ as scaling as a vacuum energy.
One can understand the unconventional scaling of $\mu$ by studying (\ref{W_flavors}); in the massless limit the
$\chi$ fields should be assigned dimension $3/4$, and hence in a perturbation of that theory $\mu$ has dimension $3/2$. Alternatively, we can derive the same scaling from \eref{oureqn}, noting that the $z_i$ are quadratic in the chiral fields of the quiver and thus have R-charge $1$ and mass dimension $3/2$. We can use the same reasoning to show that $[\epsilon]=3/2$.

}

\item
{
We have derived a gaugino mass from an open-string loop of mesonic messengers.
This is the leading contribution due to the absence of a direct soft mass arising from the D7-probe action in the DKM solution.
The general form could have been anticipated as $\mu$ is the only $R$-breaking parameter in the (undeformed) conifold geometry.
The deformation parameter $\epsilon \ll \mu$ also carries R-charge.  Therefore,
a priori one could obtain a second contribution at order $\epsilon \, \cS$ where $\epsilon$ is the KS deformation parameter.
More precisely, such a term could look like
\beq
\delta m  \sim   \frac{\epsilon^{2/3}}{\mu^{2/3}} \times  \frac{\cS}{\mu^2 \sqrt{4 \pi \lambda_\mr{eff}}}
\eeq
This correction is suppressed by a factor of $\frac{\epsilon^{2/3}}{\mu^{2/3}}$ relative to \eref{gmass}.}

\end{itemize}

\subsection{Matter soft masses from gaugino mediation}

In gaugino mediation, the MSSM scalars are usually imagined as being separated from the SUSY-breaking in extra dimensions.  This is of course
the case with our scenario; for $\mu \gg \epsilon$ even the gaugino itself is separated, but it couples much more strongly to the SUSY-breaking
than the UV-localized matter fields for $\mu \ll r^{3/2}_{\rm UV}$.

In this regime, the direct contribution to the scalar masses will be negligible \cite{gaugino}.  Instead, they will be generated by RG running
starting at the compactification scale with only a gaugino soft mass (and vanishing scalar masses and A-terms).
Assuming a unified gaugino mass $m_{\lambda}$ below the compactification scale (which should be identified with the AdS radius
$L$), one will find scalar soft masses:
\be
\label{scalar}
m^2 \sim \alpha_{SM}~ m_{\lambda}^2 ~{\rm log}\left(L M_Z \right) ~,
\ee
with coefficients determined by the quadratic Casimirs of the gauge group representation in which any given scalar sparticle transforms. Here we consider the RG running down to a scale given by the $Z$ boson mass $M_Z$. In the paper, we have considered the simplified case of an $SU(5)$ gauge group at all scales; the previous expression generalizes in the obvious way upon breaking the GUT gauge group.

Note that the gaugino mediated contribution (\ref{scalar}) to the scalar mass squared scales as ${\cal S}^2$, and thus with the fourth power of the SUSY breaking scale. In general we should also expect a non-universal gravity mediated contribution of the form $\delta m^2 \sim \beta {\cal S} / M_{UV}^2$ with $\beta$ of order 1. We will assume that the scales can be arranged such that gravity mediation is subleading, and sufficiently small to avoid dangerous flavor violating terms.

\bigskip

\section{Geometric Constraints}
\label{sec: geometric constraints}

Our model necessarily incorporates a large number of additional matter fields charged
under the SM gauge group. Special care is therefore needed in order to avoid a Landau pole. The amount of additional RG running due to the messengers
can be controlled by choosing their mass to be sufficiently heavy. In the geometric setting, this means that the D7 branes can extend only over a rather limited
range inside the warped throat geometry. In this section we compute the messenger contribution to the RG running and
quantify the associated geometric constraint.

For definiteness, we consider $K=5$ D7 branes. To simplify the discussion, we take the number of fractional branes $M$ to be constant along the RG cascade.\footnote{
For discussions of running $M$, we refer the reader to \cite{Ouyang, Benini, Benini:2007kg, Franco}.}
For scales below $\mu$, the $SU(5)$ running is controlled by the SM matter and the meson tower. Above $\mu$, the messengers contribute to the running besides the SM matter. Since the hidden sector is a large $N$ theory, it is valid to work in the approximation in which the running above $\mu$ is entirely due to the messengers. We compute the RG evolution of the 4d $SU(5)$ gauge coupling. The beta function for the inverse squared gauge coupling $x_5=8 \pi^2/g_5^2$ is
\beq
\beta_5 = 3N_c + \frac{3}{2} \sum_i (R_i -1) ~.
\label{beta_1}
\eeq
where $i$ runs over the fundamental and antifundamental messengers $\chi$ and $\tilde{\chi}$ and $N_c=5$.\footnote{More generally, the sum over $SU(5)$ flavors might involve more than one hidden sector node.}
To simplify our analysis further, we assume $N \gg M$. This condition amounts to considering
energies much larger than the hidden sector confinement scale, $\mu \gg \epsilon$. We
assume  that the mass term in \eref{W_flavors} is a small R-symmetry breaking perturbation. Then the second term in the superpotential \eref{W_flavors} determines that $R_\chi+R_A+R_B+R_{\tilde{\chi}}=2$, which implies $R_\chi+R_{\tilde{\chi}}=1.$
Substituting into (\ref{beta_1}), we get
\beq
\beta = 3 N_c - \frac{3}{2} N_f ~,
\eeq
where the number of flavors $N_f$ is the rank of the KS node to whom the MSSM is attached.

\begin{figure}[t]
\begin{center}
\psfrag{r}[cc][][.9]{$r$}
\psfrag{e}[cc][][.9]{$\epsilon$}
\psfrag{mu}[cc][][.9]{$\mu$}
\psfrag{aSM}[cc][][.9]{$\alpha_{\rm SM}^{-1}$}
\psfrag{log(L)}[cc][][.9]{$\log \Lambda$}
\includegraphics[width=8cm]{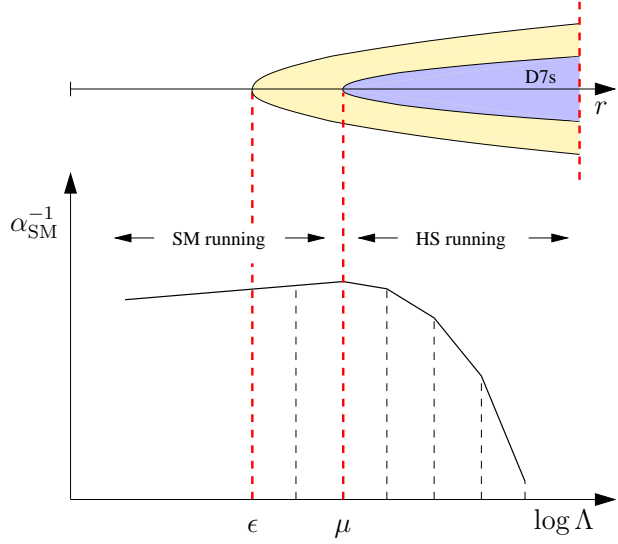}
\caption{Top: gravity dual. Bottom: running of the SM gauge coupling. Below $\mu$, the running is driven by the SM matter (SM running). Adjoint mesons can also affect the running above $\epsilon$ but, for clarity, we do not include them in the figure. Above $\mu$, the running is driven by the hidden sector (HS running), and the messengers $\chi$ and $\tilde{\chi}$ that connect the SM to the hidden sector start driving the SM
to strong coupling. The beta function is constant within each period of
the duality cascade (indicated with dashed lines).}
\label{RG_flow}
\end{center}
\end{figure}

When analyzing duality cascades, it is useful to refer to the cascade step associated with a given energy scale.
The steps in the cascade are determined via the usual expression
\beq
k_{\rm UV} - k_\Lambda  = \frac{3 g_s M}{2 \pi} \log \left(\frac{\Lambda_{\rm UV}}{\Lambda} \right) ~
\label{Lambda_k}
\eeq
A period in the cascade involves 2 dualizations. During the entire period, the beta function due to the messengers is
\beq
\beta_{k} = - \frac{3}{2} k M ~.
\eeq
\fref{RG_flow} presents a schematic picture of the RG flow.

The total variation of the squared inverse gauge coupling between $\mu$ and a scale $\Lambda$ is
\beq
\Delta x = x(\mu) - x(\Lambda) = - \sum_{k= {k}_\mu}^{{k}_\Lambda-1}  \Delta t \spc \beta_{k}  \sim \frac{3}{4} M \, \Delta t \, (k_{\Lambda}^2-k_\mu^2) ~.
\eeq
Here $\Delta t$ denotes the constant (in the probe approximation for the D7-branes)
spacing between consecutive dualizations in the cascade. We can compute $\Delta t$ to be
\beq
\label{deltat}
\Delta t =  \frac{2 \pi}{3 g_s M}
\eeq
Switching to $\alpha= 2 \pi x^{-1}$, we get
\beq
\alpha^{-1}_{\rm IR} - \alpha^{-1}_{\rm UV} \sim \frac{1}{4 g_s} (k_{\rm UV}^2 - k_{\rm IR}^2) ~.
\label{running_alpha}
\eeq

We can use this expression to constrain the length of the throat. We wish to cut-off our geometry before the scale $\Lambda_{\rm LP}$ of the Landau pole, at which $\alpha^{-1}_{\Lambda} = 0$.
We will find that $\mu$ ends up at a high scale, not too far below $\Lambda_{\rm LP}$. To get a rough estimate, suppose we take $\alpha^{-1}_\mu=25$ (a typical GUT scale value) and $g_s=1/4$. From (\ref{running_alpha}) we see that the Landau pole already occurs after just a few cascade steps: $k_{\Lambda_{\rm LP}} \lesssim 5$~.
This provides an awkwardly stringent constraint on both the length of the throat, and the depth at which the D7-branes are allowed to penetrate the throat.
Using (\ref{deltat}), and taking  $\frac{3 g_s M}{2 \pi} \sim 1$, we see that ${\mu}$ lies at most two orders of magnitude below ${\Lambda_{\rm LP}}$.

We have derived the running of the SM gauge coupling in field theory. This running has a very simple realization from a D-brane point of view as the volume spanned by the D7-branes between the corresponding radial positions.
It is not hard to show that the result \eref{running_alpha} can be reobtained by integrating the 8-d gauge coupling function $\int \sqrt{\gamma}~h$ over the 4 internal coordinates of the D7-brane between $r_{\rm IR}$ and $r_{\rm UV}$ .

The above geometric constraints on our set up are serious enough that it makes sense to try and look for ways to relax them by working with more general geometries.
One possible way of achieving this is by orbifolding. From a gravity perspective, this is a natural choice, since it reduces the volume of the D7-branes by a factor equal to the order of the discrete group. On the field theory side, the orbifold reduces the growth in the number of messenger fields.%
\footnote{A detailed analysis of cascades in orbifolds of the conifold can be found in \cite{Argurio:2008mt}.} A slight point of worry, however, is that the gain obtained by this effect is partially off-set by the fact that the volume of the $S^3$ at the tip is similarly reduced, which
could increase the minimal value of $M$ for which the anti-D3 brane configuration remains stable. We give some details of the orbifolding in appendix \ref{app: slowing down}.

\newcommand{\GUT}{{\mbox{\tiny GUT}}}

\bigskip

\section{Discussion}

We have seen that it is possible to geometrize models of strongly coupled gauge mediation using confining examples of
AdS/CFT with massive flavors.  The flavors provide messenger mesons that naturally lead to models of semi-direct gauge mediation.

The precise example we studied here has very constrained phenomenology.  The gaugino mass is
\be
m_{\lambda} \sim \frac{t}{16\pi^2} \frac{\cS}{\mu^2} (4\pi g_s N_{\rm eff})^{1/2}
\ee
(with comparable scalar soft masses arising from gaugino mediation).\footnote{This result assumes no phase cancellations and a maximum possible number of mesons $n_{\rm max}\sim (4 \pi g_s N_{\rm eff})^{1/2}$. Even if we just consider the contribution to the gaugino mass due to the the lightest meson, the estimate for $\cS^{1/4}$ is only increased by an order of magnitude.} Satisfying $m_{\lambda} \geq 100$ GeV with
moderate control parameters ( $t/16 \pi^2 \sim 10^{-2}$, $4\pi g_sN_{\rm eff} \sim 10^2$ ) requires
${\cal S} \geq 10^{3}~\mu^2 ~{\rm GeV}$ (where recall $\mu^2$ should be interpreted as having dimensions of mass$^3$).
The possibility of Landau poles forces one to consider moderately large $\mu$.
With a scale for $\mu$ of around $10^{12}$ GeV, one already has a SUSY-breaking scale $\calS^{1/4} \sim 10^{9}$-$10^{10}$ GeV, which is very close to intermediate
scale, and (in absence of sequestering) gravity mediation will be a competitive mechanism.  While the throat geometries
studied here are good candidates for sequestering \cite{KMS}, it seems that to get a safely working model it would be
best to study the slightly more elaborate orbifold geometries briefly described in \S5, where the pressure towards large $\mu$ due to Landau poles
can be relaxed.

Overall, our scenario represents a somewhat unusual hybrid mixed version of semi-direct gauge mediation and gaugino mediation. Similar to many models of (semi-)direct gauge mediation, the gaugino mass $m_\lambda$ scales with an
anomalously large power of $F/M$ (in our case,  $F^2/M^3$ instead of $F/M$, c.f. \cite{semi-direct}\cite{Komargodski:2009jf}). Normally, this will produce a low energy spectrum in which the gaugino ends up being much lighter than the scalars. In our set-up, this effect is compensated via the gaugino mediation mechanism, or equivalently, via the presence of a large number of messengers, which enhances the gaugino mass by a relative factor of $\sqrt{N}$. 

The models studied here have very few parameters, but one obvious multiparameter extension would be to allow the
locations of the chiral matter multiplets to vary as well.  Moving some of the generations in from $r = r_{\rm UV}$ to
live deeper in the throat geometry on the gravity side, corresponds to having
some of the Standard Model particles emerge as composites
during the RG cascade in the dual field theory.
SUSY-breaking models with composite generations have been studied in \cite{singlesector} and their gravity duals have
been discussed, at a phenomenological level, in \cite{GGG}.  The full understanding of gauge/gravity duality with such
composites (including the dynamics of the RG cascade and the proper computation of composite soft masses) is somewhat
involved, and our results in this direction will appear in a companion paper \cite{toappear}.
It would also be interesting to combine the kinds of ideas and techniques discussed in this paper, with
the new ideas about constructing GUT models in F-theory (for a recent review, see \cite{Vafa}). 

\bigskip
\centerline{\bf{Acknowledgements}}
\medskip
 S.F. would like to thank A. Uranga for useful conversations. H.V. acknowledges helpful discussions with M. Buican, G. Shiu, Z. Komargodski, D. Malyshev, N. Seiberg, D. Shih, T. Volansky, and B. Wecht. S.K. thanks S. Dimopoulos, T. Gherghetta, D. Green, L. McAllister, M. Mulligan, Y. Nomura, G. Shiu and J. Wacker for interesting discussions about related subjects over an extended period
of time. A.D., S.F., S.K. and H.V. acknowledge the hospitality of the Institute for Advanced Study while this work was in progress. 
S.F. and S.K. are also grateful to the Aspen Center for Physics and S.K. to the Kavli Institute for Theoretical Physics.
A.D., S.K. and D.S. are supported by the Stanford Institute for Theoretical Physics, the NSF under grant
PHY-0244728, and the DOE under contract DE-AC03-76SF00515.
The research of A.D. is also supported in part by grant RFBR 07-02-00878, and Grant for Support of Scientific Schools NSh-3035.2008.2.  The research of D.S. is also supported by the Mayfield Stanford Graduate Fellowship.
S.F. is supported by the National Science Foundation under Grant No. PHY05-51164.
F.B. is supported by the US Department of Energy under grant No. DE-FG02-91ER40671.
The research of H.V. is supported by the National Science Foundation under Grant No. PHY-0756966  and by an Einstein Fellowship of the Institute for Advanced Study.

\newpage

\appendix

\section{The conifold and a normalizable non-SUSY deformation}
\label{app: sugra solutions}

We start introducing some conventions about the conifold geometry.
The unwarped metric on the singular conifold is
\be
ds_6^2 = dr^2 + r^2 \Big\{ \sum_{i=1,2} \big( (e^{\theta_i})^2 + (e^{\varphi_i})^2 \big) + (e^\psi)^2 \Big\} \;,
\ee
where we defined the following 1-forms on $T^{1,1}$:
\be
e^{\theta_i} = \frac{1}{\sqrt 6} \, d\theta_i \;, \qquad e^{\varphi_i} = \frac{1}{\sqrt 6} \, \sin\theta_i\, d\varphi_i \;, \qquad e^\psi = \frac{1}{3} \Big( d\psi - \sum_{i=1,2} \cos\theta_i\, d\varphi_i \Big) \;.
\ee
The unwarped K\"ahler form and holomorphic form are given by
\begin{align}
J &= r\, dr \wedge e^\psi + \sum_{i=1,2} r^2\, e^{\theta_i} \wedge e^{\varphi_i} \\
\Omega &= \frac{9}{4}\, r^2 e^{i\psi} \big( dr + i\, r\, e^\psi \big) \wedge \big( e^{\theta_1} + i\, e^{\varphi_1} \big) \wedge \big( e^{\theta_2} + i\, e^{\varphi_2} \big) \;. \label{omega3}
\end{align}
In fact $dJ = d\Omega = 0$. We can introduce holomorphic coordinates which make explicit the $SO(4)$ symmetry of the geometry:
\bea
z_1 &= \frac{1}{\sqrt 2} \, r^{3/2} e^{i\psi/2} \Big\{ e^{-\frac{i}{2}(\varphi_1 + \varphi_2)} \sin\frac{\theta_1}{2} \sin\frac{\theta_2}{2} + e^{\frac{i}{2}(\varphi_1 + \varphi_2)} \cos\frac{\theta_1}{2} \cos\frac{\theta_2}{2} \Big\}\\
z_2 &= \frac{1}{\sqrt 2 \,i} \, r^{3/2} e^{i\psi/2} \Big\{ e^{-\frac{i}{2}(\varphi_1 + \varphi_2)} \sin\frac{\theta_1}{2} \sin\frac{\theta_2}{2} - e^{\frac{i}{2}(\varphi_1 + \varphi_2)} \cos\frac{\theta_1}{2} \cos\frac{\theta_2}{2} \Big\}\\
z_3 &= \frac{1}{\sqrt 2} \, r^{3/2} e^{i\psi/2} \Big\{ e^{-\frac{i}{2}(\varphi_1 - \varphi_2)} \sin\frac{\theta_1}{2} \sin\frac{\theta_2}{2} - e^{\frac{i}{2}(\varphi_1 - \varphi_2)} \cos\frac{\theta_1}{2} \cos\frac{\theta_2}{2} \Big\}\\
z_4 &= \frac{1}{\sqrt 2 \,i} \, r^{3/2} e^{i\psi/2} \Big\{ e^{-\frac{i}{2}(\varphi_1 - \varphi_2)} \sin\frac{\theta_1}{2} \sin\frac{\theta_2}{2} + e^{\frac{i}{2}(\varphi_1 - \varphi_2)} \cos\frac{\theta_1}{2} \cos\frac{\theta_2}{2} \Big\} \;.
\eea
The conifold equation reads $\sum_{i=1}^4 z_i^2 = 0$, and the radial coordinate is $r^3 = \sum_{i=1}^4 |z_i|^2$. In terms of such coordinates we can construct a basis of $SO(4)$-invariant $(1,1)$ forms:
\bea
i\, dz_j \wedge d\bar z_j &= \frac{9}{2}\, r^2\, dr \wedge e^\psi + 3r^3 {\textstyle \,\sum_{i=1,2}\,} e^{\theta_i} \wedge e^{\varphi_i} \\
i(\bar z_j\, dz_j) \wedge (z_k\, d\bar z_k) &= \frac{9}{2}\, r^5 \, dr \wedge e^\psi \\
\lambda_2 \equiv i\, \epsilon_{ijkl}\, z_i \bar z_j \, dz_k \wedge d\bar z_l &= 3 r^6 \big( e^{\theta_1} \wedge e^{\varphi_1} - e^{\theta_2} \wedge e^{\varphi_2} \big) \;.
\eea

The Klebanov-Tseytlin (KT) supergravity solution \cite{KT} describes the supersymmetric configuration of $N$ regular and $M$ fractional D3-branes at the tip of the conifold, at least at radii larger than the gaugino condensation scale: $r > \epsilon^{2/3} \sim \alpha' \big( \langle \lambda\lambda \rangle / M \big) \phantom{}^{1/3}$; this is because such solution does not include confinement and chiral symmetry breaking. It is specified by:
\bea
ds_{10}^2 &= h(r)^{-1/2} \, \eta_{\mu\nu} dx^\mu dx^\nu + h(r)^{1/2} ds_\mathrm{conifold}^2 \\
g_s F_5 &= (1 + *) \, dh^{-1} \wedge \dvol_{3,1} \\
B_2 &= \frac{3 g_s M \alpha'}{2} \, \log\frac{r}{r_0} \, \omega_2 \equiv \frac{k(r)}{3} \, \omega_2 \qquad\qquad \omega_2 = 3 \big( e^{\theta_1} \wedge e^{\varphi_1} - e^{\theta_2} \wedge e^{\varphi_2} \big) \;,
\eea
moreover $H_3 = dB_2$, $F_3 = e^{-\Phi} *_6 H_3$, the axio-dilaton is constant and the warp factor is
\be
h(r) = \frac{27 \pi \alpha'^2}{4r^4} \Big[ g_s N + \frac{3(g_sM)^2}{2\pi} \Big( \frac{1}{4} + \log\frac{r}{r_0} \Big) \Big]\;.
\label{warp_factor_KT}
\ee
Here $r_0$ is the scale where the running D3-charge is $N$. If we tune it to the IR such that $N=0$, then it can be identified with the gaugino condensate scale: $r_0 \simeq \epsilon^{2/3}$.
We can rewrite the unwarped K\"ahler form and the B-field in an $SO(4)$-invariant way:
\begin{align}
J &= \frac{i}{3r} \, dw_j \wedge d\bar w_j - \frac{i}{9r^4} \, (\bar w_j\, dw_j) \wedge (w_k\, d\bar w_k ) \label{J in KT as symmetric}\\
B_2 &= \frac{k(r)}{3r^6} \, i \, \epsilon_{ijkl} \, w_i\, \bar w_j \, dw_k \wedge d\bar w_l \;.  \label{B in KT as symmetric}
\end{align}

\subsection{Non-SUSY deformation of the conifold}

In order to describe the SUSY-breaking vacuum from a gravity perspective, it is necessary to understand the backreaction of $\overline{{\rm D3}}$-branes at the tip of the KS geometry.
The DKM solution provides the first order (in the vacuum energy) perturbation to the KT asymptotic UV background due to the anti-branes
\cite{DKM}. To simplify the calculation, DKM considered the case in which the $\overline{{\rm D3}}$-branes are smeared over the internal space. In the field theory, this corresponds to integrating over Goldstone modes around the vacuum. As a result, the background preserves Poincar\'e invariance and all the symmetries of $T^{1,1}$, namely $SU(2)\times SU(2)\times \mathbb{Z}_2$. In addition, $U(1)_\psi$ is also preserved.
The metric has the form
\beq
ds^2 = h(r)^{-1/2} \eta_{\mu\nu} dx^\mu dx^\nu + h(r)^{1/2} \Big( dr^2 + r^2 \sum_{i=1,2} \big( (e^{\theta_i})^2 + (e^{\varphi_i})^2 \big) + r^2 e^{2b(r)} (e^\psi)^2 \Big) \;.
\eeq
The rest of the fields are given by
\bea
F_3 &= \bar{M}\, e^\psi \wedge (e^{\theta_1} \wedge e^{\phi_1} - e^{\theta_2} \wedge e^{\phi_2}), & \qquad
B_2 &= k(r) (e^{\theta_1} \wedge e^{\phi_1} - e^{\theta_2} \wedge e^{\phi_2}) , \\
\tilde{F}_5 &= dC_4 - C_2 \wedge H_3, &
g_s C_4 &= \alpha(r) \, \dvol_{3,1}, \\
\Phi &= \Phi(r), & C_0 &= 0 \;.
\eea
The solution, with the first corrections in $\cS$ and $1/r^4$, is given in Einstein frame by:
\begin{align}
\frac{r^4}{\alpha'^2}\, h(r) &= \frac{1}{4}\, g_s \bar{N} + \frac{1}{8} (g_s \bar{M})^2 + \frac{1}{2} (g_s \bar{M})^2 \log\frac{r}{r_0} \nonumber \\
& \quad + \frac{\alpha'^4}{r^4} \Big[ \Big( \frac{1}{32}\, g_s \bar{N} + \frac{13}{64} (g_s \bar{M})^2 + \frac{1}{4} (g_s \bar{M})^2 \log\frac{r}{r_0} \Big) \cS - \frac{1}{16} (g_s \bar{M})^2 \phi \Big] \\
e^{2b(r)} &=  1 + \frac{\alpha'^4\cS}{r^4} \\
\frac{1}{\alpha'}\, k(r) &= g_s \bar{M} \log\frac{r}{r_0} + \frac{\alpha'^4}{r^4} \Big[ \Big( \frac{3}{8} \frac{\bar N}{\bar M} + \frac{11}{16} \, g_s \bar{M} + \frac{3}{2}\, g_s \bar{M} \log\frac{r}{r_0} \Big) \cS - \frac{1}{4}\, g_s \bar{M} \phi \Big] \\
\Phi(r) &= \log g_s + \frac{\alpha'^4}{r^4}\,  \big[ \phi - 3 {\cal S} \log\frac{r}{r_0} \big] \\
\alpha(r) &= h(r)^{-1} \;.
\end{align}
Here $\bar{N}=27 \pi N$, $\bar{M} = \frac{9}{2} M$. ${\cal S}$ is the vacuum energy of the metastable
vacuum
\be
{\cal S} \sim \frac{p}{N} \, e^{- 8\pi N /(3 g_s M^2)} \, \frac{r_0^4}{\alpha'^4} \;.
\ee
We will not be interested in the perturbation $\phi$ and will set $\phi =0$.

The backreaction of the $\overline{{\rm D3}}$-branes introduces IASD flux, squashes the internal space while preserving its isometries (with relative warping between the two factors of $T^{1,1}$ given by $e^b$) and makes the dilaton run.

As opposed to other non-SUSY modifications of the KS solution \cite{Kuperstein:2003yt}, DKM corresponds to a {\it normalizable} perturbation. Because of that, it
is dual to {\it spontaneous}, as opposed to explicit, SUSY-breaking.

\section{Dirac action on the D7-brane}
\label{DiracSection}

Our goal is to derive the expression for the tree-level gaugino mass
and to show that it vanishes in the case of DKM background. Our starting point is the
general action for fermions derived in \cite{Martucci}. At the lowest level in perturbation theory in $\cal S$
there is no need
to diagonalize the mass matrix. Rather one can use the gaugino wave function $\lambda$ defined in the SUSY case and calculate the coefficient in front
of $Tr(\lambda^2)$ up to ${\cal O}(\cal S)$.
As we discussed in section \ref{S4.1}, when SUSY is not broken the gaugino can be defined as a scalar invariant under the action of ``holomorphic'' $SU(3)$ \cite{CIU}.
Hence, we are only interested in collecting the terms proportional to
\bea
\label{123}
\bar{\theta}\Gamma_{\bar{1}\bar{2}\bar{3}}\theta = Tr(\lambda^2)\ ,\\
\bar{\theta}\Gamma_{123}\theta = Tr({\bar \lambda}^2)\  ,
\eea
where $1,2,3$ are the holomorphic indexes along the six internal dimensions of the conifold. Clearly (\ref{123}) is invariant under the holomorphic change of variables.

Since $\theta$ is a ten-dimensional Majorana-Weyl spinor only
terms of the form $\bar{\theta}\Gamma_{n_1..n_k}\theta$
with $k=3$ or $7$ are non-zero. In a given point we can always choose the directions $1,2,{\bar 1},{\bar 2}$
to be aligned with the D7-brane and the directions $3,{\bar 3}$ to be orthogonal.
Hence
\bea
\bar{\theta} \check{\Gamma}_7^{-1}\Gamma_{a_1..a_k}\theta=-i\bar{\theta} \Gamma_{3\bar{3}}\Gamma_{a_1..a_k}\theta\ .
\eea
The coupling $F_5^{n_1..n_k}\bar{\theta} \check{\Gamma}_7^{-1}\Gamma_{n_1..n_k}\theta$ therefore does not contribute to the gaugino mass and the only relevant terms are (\ref{123}) and 
\bea
\bar{\theta} \check{\Gamma}_7^{-1}\Gamma_{{\bar 1}{\bar 2} {\bar 3}}\theta=iTr(\lambda^2)\ ,\\
\label{m1}
\bar{\theta} \check{\Gamma}_7^{-1}\Gamma_{123}\theta=-iTr(\bar{\lambda}^2)\ .
\eea
The minus sign in (\ref{m1}) is crucial in ensuring that the gaugino mass is real.

Before we proceed with analyzing the Dirac action in detail let us introduce the ``Fierz brackets'' $[\ \ ]$ which will denote the $\Gamma_{\bar{1}\bar{2}\bar{3}}$ component of the gamma-matrices product
\bea
\Gamma_{n_1}..\Gamma_{n_k}=[\Gamma_{n_1}..\Gamma_{n_k}]\Gamma_{\bar{1}\bar{2}\bar{3}}+...
\eea
The only terms relevant for the gaugino mass  are
\bea
\label{mass1}
{1\over 2!4}[\Gamma_i \Gamma_{nm} ] (M^{-1})^{ij}g^{np}g^{mq} H_{j pq}\ ,
\eea
\bea
\label{mass2}
-{1\over 3!4}[\Gamma_{nmk} ] g^{np}g^{mq}g^{kl} H_{pql}\ ,
\eea
and
\bea
\label{mass3}
{-i\over 3!8}[\Gamma_{i}\Gamma_{nmk}\Gamma_{j}] (M^{-1})^{ji}g^{np}g^{mq}g^{kl} e^{\Phi}F_{pql}\ ,
\eea
\bea
\label{mass4}
{-i\over 3!4}[\Gamma_{nmk}] g^{np}g^{mq}g^{kl} e^{\Phi}F_{pql}\ .
\eea

In the first term (\ref{mass1}) only $\alpha\neq n,m$ contribute leading in the SUSY case to
\bea
\label{mass1_1}
{1\over 2!4}[ \Gamma_{i nm} ] (M^{-1})^{ij} g^{np}g^{mq} H_{j pq}=\\
{ (M^{-1})^{ij}\gamma_{ij} \over 3!8}[ \Gamma_{nmk} ] g^{np}g^{mq}g^{kl} H_{pql}\ .
\eea
This calculation can be easily done in the special coordinates in which $g_{a\bar b}$ is diagonal. There is no need at this point to align the coordinates $1,2$ along the D7-brane.
Neglecting the world-volume flux, we have ${(M^{-1})^{\alpha\beta}\gamma_{\alpha\beta} \over 3!8}={2\over 3! 4}$. Essentially, this is a sign change in front of (\ref{mass2}).

In the SUSY case when $M_{ij}$ is of $(1,1)$ type the last term (\ref{mass4}) simply does not contribute to (\ref{m1}).

Using the definition of the 3-form flux
\bea
G^3=F-ie^{-\Phi}H\ ,
\eea
and the holomorphic indexes
one can express $F$ and $H$ through $G^3$
\bea
2F_{abc}=G^3_{abc}+(G^3_{\bar{a}\bar{b}\bar{c}})^*\ ,\\
2ie^{-\Phi}H_{abc}=(G^3_{\bar{a}\bar{b}\bar{c}})^*-G^3_{abc}\ .
\eea
In this notation, the gaugino mass term in the SUSY case is proportional to
\bea
\label{msusy}
{g_s\over 2} \left( \left((G^3_{{\bar 1}{\bar 2}{\bar 3}})^*-G^3_{123}\right)+\left(G^3_{123}+(G^3_{{\bar 1}{\bar 2}{\bar 3}})^*\right)\right)Tr(\lambda^2) +c.c. =
g_s(G^3_{{\bar 1}{\bar 2}{\bar 3}})^* Tr(\lambda^2)+c.c.
\eea
Hence we have reproduced the result of \cite{CIU} that the gaugino mass is sourced only by $(0,3)$ flux
in the SUSY case when no world-volume flux is present.

If there is non-trivial world-volume flux, (\ref{mass1_1}) produces an additional term
\bea
{ Tr(M^{-1}-\gamma^{-1}) }[ \Gamma_{\bar a \bar b\bar c} ] g^{\bar a d}g^{\bar b e}g^{\bar c e}g_s\left( (G^3_{\bar{d}\bar{e}\bar{f}})^*-G^3_{def}\right) \, ,
\eea
which contains both $(0,3)$ and $(3,0)$ flux. Clearly, this term is also zero in the KS case.

In the non-SUSY case of the DKM background we need to perturb the fluxes and metric up to the linear order in ${\cal S}$.
Since the perturbed flux $\delta G^3_{abc}$ does not contain the $(0,3)$ or $(3,0)$ piece the perturbation of the 3-form flux can be neglected.

In addition to the $(1,1)$ piece the perturbation of the background metric $\delta g^{mn}$ will have the $(2,0)$ and $(0,2)$ pieces. The $(1,1)$ part is not ``dangerous''
as it couples the gaugino mass to the absent $(0,3)$ and $(3,0)$ flux.  On the contrary, the
$(0,2)$ piece in (\ref{mass2}) and (\ref{mass4}) will lead to
\bea
\label{dm1}
 [\Gamma_{\bar{a}\bar{b}\bar{c}}]\delta g^{\bar{a}\bar{d}}g^{\bar{b}e}g^{\bar{c}f} g_s G^3_{\bar{d}ef}\ ,
\eea
i.e. will couple the gaugino mass to the $(2,1)$ flux.
Similarly the perturbation of (\ref{mass1}) gives
\bea
\label{dm2}
 [\Gamma_{\bar{\alpha}\bar{a}\bar{b}}](M^{-1})^{\bar{\alpha}\beta}\delta g^{\bar{a}\bar{c}}g^{\bar b d} g_s \left((G^3_{\bar{\beta} c \bar{d}})^*-G^3_{\beta \bar{c}d}\right)\ ,
\eea
while the perturbation of (\ref{mass4}) with respect to $g^{mn}$ vanishes.

Both terms (\ref{dm1}) and (\ref{dm2}) are potentially dangerous as they couple the gaugino mass to the non-zero $(2,1)$ flux. Nevertheless they both vanish in the particular case of the DKM solution because in this case
\bea
\delta g^{\bar a \bar b}G^3_{\bar b c d}=\delta g^{\bar a \bar b}(G^3_{ b \bar c \bar d})^*=0\ .
\eea
Indeed using the notations of appendix \ref{app: sugra solutions} we can introduce the basis of the holomorphic 1-forms
\bea
Z_i=d\theta_i-i\sin\theta_i d\phi_i\  \ \ i=1,2\ ,\\
Z_3={dr}+ i \, r \, e^\psi \ .
\eea
The coordinates $Z_i$ diagonalize the metric but are not special from the D7-brane's embedding point of view.
In this notation, the only non-trivial 3-flux is the $(2,1)$ form
\bea
\label{g3}
G^3\sim (g_S M \alpha')r^{-1} Z_3\wedge (Z_1\wedge {\bar Z_1}-Z_2\wedge {\bar Z_2})\ ,
\eea
and the only non-trivial $(0,2)$ perturbation of the background metric is $\delta g^{{\bar 3}{\bar 3}}$. Therefore the combination of unperturbed $G^3_{a b \bar c}$ or $(G^3_{\bar a\bar b c})^*$ and  $\delta g^{\bar{a}\bar{b}}$ vanish. Let us note that the situation would be quite different if a non-trivial $\delta g^{{\bar a }{\bar b}}$ for $a=1,2$ was present.  However, that would be inconsistent with the $SU(2)\times SU(2)$ symmetry of the DKM solution.

The last ingredient is the perturbation of the combination of the induced metric and world-volume flux $M^{-1}$.
Again, the perturbation of the $(1,1)$ piece does not affect the SUSY result that only $(0,3)$ and $(3,0)$ flux contribute.
But the presence of the $(2,0)$ and $(0,2)$ pieces changes that.
Thus the contribution (\ref{mass3}) is not zero anymore. Rather
\bea
\label{mass3_1}
{1\over 3!}[\Gamma_{i}\Gamma_{nmk}\Gamma_{j}] \delta (M^{-1})^{ji}g^{np}g^{mq}g^{kl} F_{pql}=
{2\over 2!}[\Gamma_{\bar \alpha  \bar a\bar b}] \delta (M^{-1})^{(\alpha\beta)} g^{\bar a c}g^{\bar b d} F_{\bar \beta c d}\ .
\eea
It is easy to check this identity in the special coordinates in which the metric $g_{mn}$ is diagonal.
Together with the perturbation of (\ref{mass1}) this combines into
\bea
\label{dgmass}
[\Gamma_{{\bar \alpha} {\bar a}{\bar b}}]\delta (M^{-1})^{({\bar \alpha}{\bar \beta} )}g^{{\bar a}c}g^{{\bar b}d}g_s(G^3_{ \beta { \bar c}{\bar d}})^*\ ,
\eea
and
\bea
\label{dgmass_assym}
[\Gamma_{{\bar \alpha} {\bar a}{\bar b}}]\delta (M^{-1})^{[{\bar \alpha}{\bar \beta} ]}g^{{\bar a}c}g^{{\bar b}d}g_s\left((G^3_{ \beta { \bar c}{\bar d}})^*-G^3_{\bar \beta c d} \right) \ .
\eea
Hence whenever the induced metric contains a $(0,2)$ piece, the gaugino couples to the IASD $(1,2)$ flux through (\ref{dgmass})
and to the ISD $(1,2)$ and IASD $(2,1)$ flux through (\ref{dgmass_assym}). Since in the KS solution only the ISD $(2,1)$ flux is present all couplings above vanish.

It is important to note that the interplay of signs was crucial in obtaining the zero couplings above. In particular the vanishing coupling of the gaugino to the IASD $(1,2)$ flux (\ref{dgmass})
could easily be turned into the coupling to the ISD $(2,1)$ flux
\bea
[\Gamma_{{\bar \alpha} {\bar a}{\bar b}}]\delta \gamma^{{\bar \alpha}{\bar \beta} }g^{{\bar a}c}g^{{\bar b}d}G^3_{ {\bar \beta} c d}
\eea
by a flip of sign (here we dropped the world-volume flux for simplicity). Clearly this term is naively expected to be present and would lead to a non-zero mass. To estimate it we would have to calculate the ratio of the integrals over the internal directions of the D7-brane
\bea
\label{ratiom}
m={\int d^4 x \int  d^4 y \sqrt{\gamma}\  \delta \gamma^{{\bar \alpha}{\bar \beta} }g^{{\bar a}c}g^{{\bar b}d}g_s G^3_{ {\bar \beta} c d}\ \bar{\theta}\ \Gamma_{{\bar \alpha} {\bar a}{\bar b}}\ \theta \over \int d^4 x \int d^4 y \sqrt{\gamma}\ \bar{\theta}\ \Gamma^{x_0} \theta\ }\ .
\eea
Here we assume that only gaugino mode of $\lambda$ is turned on within $\theta$.
The result of \cite{MMS} for the scaling of the gaugino wave function
\bea
\theta\sim h^{3/8}\ ,
\eea
together with the scaling of the  gamma-matrices $\Gamma^{x_0}\sim h^{1/4}$ suggests
the denominator to be proportional to the warped volume of the D7-brane i.e. to the inverse coupling constant of the world-volume gauge theory 
\bea
\int d^4 y \sqrt{\gamma} h \sim g^{-2} L^4\ ,\ \ \  L^2=(g_s M \alpha')\, .
\eea

The only nontrivial relevant combination from the numerator of (\ref{ratiom}) is
\bea
(g_s M \alpha') \gamma^{{\bar 1} {\bar 2}} g^{{\bar 3}{\bar 3}} \left(g^{{\bar 1}{\bar 1}}+g^{{\bar 2}{\bar 2}}\right)\Gamma_{{\bar 1}{\bar 2}{\bar 3}}\ .
\eea
To calculate this expression we need the scaling of the gamma-matrices along the internal directions
(compare with (\ref{omega3}))
\bea
\Gamma_{{\bar 1}{\bar 2}{\bar 3}} \sim
h^{3/4} e^{-i\psi}r^2 {\bar Z}_1\wedge {\bar Z}_2\wedge {\bar Z}_3\ ,
\eea
as well as the perturbed induced metric
\bea
\delta \gamma^{{\bar 1}{\bar 2}}={36\over  h^{1/2} r^2}{(r^3-|\mu|^2)\over (4r^3-|\mu|^2)^2}{\bar \mu^2} e^{i\psi} {{\cal S}\over r^4}\ .
\eea
All terms together lead to
\bea
m\sim g_s(g^{-2}L^4)^{-1} (g_s M\alpha')\bar \mu^2 {\cal S}\int_{r^3=|\mu|^2}^{r_{cutoff}} {dr\over r^7} {(r^3-|\mu^2|)\over 4r^3-|\mu^2|}\sim {g_s g^2{\cal S}\over \mu^2 L^2}={t{\cal S}\over K \mu^2 L^2}\ ,
\eea
which coincides with the one-loop calculation (\ref{gmass}) in the case of one probe D7-brane.

\section{Bosonic action on the D7-brane}
\label{app: effective action}


We can embed D7-branes in the singular (KT solution) or deformed (KS solution) conifold along the so-called massive Kuperstein embedding, given by the equation $z_4 = \mu$. It is a curve described by the equation $z_3^2 = \epsilon^2 - \mu^2 - z_1^2 - z_2^2$ (possibly with $\epsilon=0$), which represents a deformed $\bbC^2/\bbZ_2$ geometry. We can parametrize it locally with $(z_1, z_2)$ treating $z_3$ as a function. After the substitution
\be \label{holomorphic pullback}
dz_3 = - \frac{z_1\, dz_1 + z_2\, dz_2}{z_3} \;,
\ee
it is straightforward to compute from (\ref{J in KT as symmetric})-(\ref{B in KT as symmetric}) that on the embedding:
\be
\hat B_2 \wedge \hat J^{(1)} = 0 \qquad\qquad\qquad \hat B_2 \wedge \hat J^{(2)} = 0 \;,
\ee
where $J^{(i)}$ are the two pieces appearing in (\ref{J in KT as symmetric}) (similar expressions for $B_2$ and $J$ with only different radial functions hold in the KS case, see (\ref{KS solution})).
Therefore setting $\calF = \hat B_2$ on the D7-brane we solve the following equations:
\be \label{flux SUSY equations}
\calF \text{ is } (1,1) \;, \qquad\qquad\quad \calF \wedge \hat J = 0 \;, \qquad\qquad\quad d\calF = \hat H_3 \;.
\ee
These, together with the requirement that the embedding is holomorphic, are the conditions for the D7-brane with worldvolume flux $\calF$ to be supersymmetric in our setup \cite{Becker:1995kb, Marino:1999af, Gomis:2005wc}.

We can even add an extra worldvolume flux $F_2$ which wraps the 2-cycle of $\bbC^2/\bbZ_2$. In the singular conifold case, setting $\calF = \hat B_2 + 2\pi\alpha' F_2$ and
\be \label{KT worldvolume flux}
2\pi\alpha' F_2 = \frac{P}{2r^6} \, 2\re \big[ i \bar\mu \, \epsilon_{ijk} \, z_i \, dz_j \wedge d\bar z_k \big] \Big|_{z_4 = \mu}
\ee
solves again the SUSY equations (\ref{flux SUSY equations}); $P \in \bbZ$ because of the quantization condition $\int_{S^2} F_2 = 2\pi P$. A similar expression is valid in the deformed conifold case.

In the following we will be mainly interested in the singular conifold. We can rewrite it using a set of coordinates particularly useful for the Kuperstein embedding and its oscillations. First of all we parametrize the holomorphic coordinates as:
\bea \label{holomorphic coordinates wp}
z_1 &= i (\mu + \chi) \Big[ \cos\phi\, \cosh \Big( \frac{\rho + i\gamma}{2} \Big) \cos\theta - i \sin\phi\, \sinh \Big( \frac{\rho + i\gamma}{2} \Big) \Big] \\
z_2 &= i (\mu + \chi) \Big[ \sin\phi\, \cosh \Big( \frac{\rho + i\gamma}{2} \Big) \cos\theta + i \cos\phi\, \sinh \Big( \frac{\rho + i\gamma}{2} \Big) \Big] \\
z_3 &= i (\mu + \chi) \cosh \Big( \frac{\rho + i\gamma}{2} \Big) \sin\theta \\
z_4 &= \mu + \chi \;,
\eea
where $\chi$ is a complex variable, whilst $\rho$, $\gamma$, $\theta$, $\phi$ are real. The range of coordinates is: $\chi,\bar\chi \in \bbC$, $\rho \in [0,\infty)$, $\gamma \in [0, 4\pi)$, $\theta \in [0,\pi]$, $\phi \in [0,2\pi)$. The massive Kuperstein embedding is easily described by $\chi = 0$; $\chi$ and $\bar \chi$ are then our transverse coordinates, and in fact we have written the singular conifold as a foliation of Kuperstein embeddings. Then
\be
r^3 = \sum_{i=1}^4 |z_i|^2 = |\mu + \chi|^2 ( \cosh\rho + 1) \;.
\ee
To exploit the $SO(3)$-invariance of the embedding, we introduce left-invariant forms:
\bea \label{left-invariant forms}
h_1 &= 2 \Big( \cos\frac{\gamma}{2}\, d\theta - \sin\frac{\gamma}{2}\, \sin\theta\, d\phi \Big) \\
h_2 &= 2 \Big( \sin\frac{\gamma}{2}\, d\theta + \cos\frac{\gamma}{2}\, \sin\theta\, d\phi \Big) \\
h_3 &= d\gamma - 2\cos\theta\, d\phi \;.
\eea
They satisfy the Maurer-Cartan equations $dh_i = \frac{1}{4} \epsilon_{ijk} h_j \wedge h_k $ for $i,j,k=1,2,3$; moreover $\int_{S^3/\bZ_2} h_1 \wedge h_2 \wedge h_3 = 64\pi^2$.

We can finally rewrite the DKM metric and B-field as:
\bea \label{DKM metric B-field invariant}
ds_6^2 &= \frac{1}{3r} \bigg\{ \frac{|\mu + \chi|^2}{2} \Big[ \frac{1 + 2\cosh\rho}{3}\, \big( d\rho^2 + h_3^2 \big) + \cosh^2\frac{\rho}{2}\, h_1^2 + \sinh^2\frac{\rho}{2}\, h_2^2 \Big] \\
&\qquad + \frac{2}{3}\, \sinh\rho\, (d\rho + i\, h_3) \, (\mu + \chi)\, d\bar\chi + \mathrm{c.c.} + \frac{4}{3}\, (1+\cosh\rho) \, d\chi\, d\bar\chi \\
&\qquad + \frac{\calS}{r^4} \bigg[ |\mu + \chi|^2 \, \frac{\cosh\rho -1}{3} \, h_3^2 + \frac{1 + \cosh\rho}{3} \Big( 2 \, d\chi d\bar\chi - \frac{\bar\mu + \bar\chi}{\mu + \chi} \, d\chi^2 - \frac{\mu + \chi}{\bar\mu + \bar\chi} \, d\bar\chi^2 \Big) \\
&\qquad\qquad + \frac{2i}{3} \sinh\rho\, h_3\, (\mu + \chi) \, d\bar\chi + \mathrm{c.c.} \bigg] \bigg\} \\
B_2 &= \frac{k(r)}{3r^6} \, |\mu + \chi|^4 \, \cosh^2\frac{\rho}{2}\, \Big( - \sinh\frac{\rho}{2}\, d\rho \wedge h_2 + \cosh\frac{\rho}{2}\, h_3 \wedge h_1 \Big) \;.
\eea
Notice that the pulled-back metric is diagonal, and $B_2$ does not have components along the transverse directions.

\subsection{The D7-brane effective action}

We compute the induced action on the Kuperstein D7-brane in the DKM solution. We use the following conventions for indices: $a,b$ for the eight D7 directions, $i,j$ for the two transverse directions, $\alpha,\beta$ for the four internal D7 directions. $\phi^i = \{ \chi,\, \bar\chi\}$ are the transverse scalar fields and $A_a$ is the worldvolume gauge field. We absorb $2\pi\alpha'$ into $F_2$ and set $\alpha'=1$ unless explicitly reintroduced.

The Einstein frame bosonic action is:
\be
S_{D7}^\mathrm{bosonic} = -\frac{\mu_7}{g_s^2} \int d^8\xi\, e^{\Phi} \sqrt{ -\det \big( \hat g_{ab} + g_s^{1/2} e^{-\Phi/2} \cF \big)} \, + \mu_7 \int \frac{1}{2}\, C_4 \wedge \cF^2 \;,
\ee
where $\cF = \hat B_2 + F_2^\mathrm{bg} + F_2$ and $\mu_7 = g_s (2\pi)^{-7}$. We expand it up to quadratic terms in the worldvolume fields, and after integration on the $SO(3)$ isometry we get:
\bea
\label{reduced action}
S &= - \frac{1}{2\pi^5 g_s} \,  \int d^4x\,d\rho\, \big( \cL_A + \cL_B + \cL_C \big) \\
\cL_A &= - \frac{e^\Phi \sqrt{-\tilde q}}{4} \Big\{ \tilde q^{(ab)}  F_{bc}  \tilde q^{(cd)} F_{da}
+ \, \tilde q^{[\alpha\beta]} F_{\beta\gamma}  \tilde q^{[\gamma\delta]}  F_{\delta\alpha} - \frac{1}{2} \bigl(\tilde q^{[\alpha\beta]}  F_{\beta\alpha} \bigr)^2 \Big\} + \frac{1}{8h} \, F_{\alpha\beta} F_{\gamma\delta} \, \epsilon^{\alpha\beta\gamma\delta} \\
\cL_B &= \frac{e^\Phi \sqrt{-\tilde q}}{2} \Big\{ \tilde q^{(ab)} q_{ij} \partial_a \phi^i \partial_b \phi^j +
 \tilde q^{\rho\rho} q_{i\rho}\partial_\rho \phi^i \Big( \tilde q^{\rho\rho} q_{j\rho}\partial_\rho \phi^j
+  \tilde q^{[\alpha\beta]}  F_{\beta\alpha} \Big)  \\
& \qquad\qquad\qquad\qquad\qquad\qquad  - 2\, \tilde q^{(ab)} q_{i(b}\partial_{c)}\phi^i  \Big( \tilde q^{(cd)} q_{j(d}\partial_{a)}\phi^j + 2 \, \tilde q^{[cd]}  \,  F_{da} \Big) \Big\}  \\
\cL_C &=  e^\Phi \sqrt{-\gamma} \, \Big|_{\leq 2} + e^\Phi \sqrt{-\tilde q} \, \tilde q^{ab} \, q_{i(a} \partial_{b)} \phi^i \, \Big|_{\leq 2} \;.
\eea
${\cal L}_A$ contains only gauge kinetic terms. ${\cal L}_B$ consists of kinetic terms for $\chi,\, \bar\chi$ and mixed kinetic terms. The third term ${\cal L}_C$ is obtained by expanding $e^\Phi$, $\gamma$, and $q$ in powers of the fields $\chi,\, \bar\chi$, and gives rise to a potential for the scalars. In the SUSY background it vanishes (up to a field-independent term that can be removed), while in the DKM background it induces a VEV for the worldvolume fields.

Here $q_{MN} = g_{MN} + {g_s}^\frac{1}{2} e^{-\frac{\Phi}{2}} \, B_{MN}$ as directly read from (\ref{DKM metric B-field invariant}), while the pulled-back quantities $\tilde q_{ab}$ and $\gamma_{ab}$ are defined as follows:
\be
\tilde q_{ab} = \hat q_{ab} + {g_s}^\frac{1}{2} e^{-\frac{\Phi}{2}} \, F^\mathrm{bg}_{ab} \,, \qquad\qquad \tilde q_{(ab)} \equiv \gamma_{ab} \,, \qquad \tilde q_{[\alpha\beta]} = {g_s}^\frac{1}{2} e^{-\frac{\Phi}{2}} \big( \hat B_{\alpha\beta} + F^\mathrm{bg}_{\alpha\beta} \big) \;.
\ee
In the basis $x^0, \dots x^3,\, \rho,\, h_3,\, h_1,\, h_2$, the actual computation gives:
\bea \label{metric gamma}
\gamma_{\alpha\beta} &= \frac{h^{1/2} |\mu + \chi|^2}{6r} \: \text{diag} \Big( \frac{1+2\cosh\rho}{3},\, \frac{1+2\cosh\rho}{3} + \frac{2 \calS}{r^4} \, \frac{\cosh\rho - 1}{3},\, \cosh^2\frac{\rho}{2},\, \sinh^2\frac{\rho}{2} \Big)_{\alpha\beta} \\
\gamma_{\mu\nu} &= h^{-1/2}\, \eta_{\mu\nu} \\
\sqrt{-\gamma} &= \frac{|\mu + \chi|^4}{36\, r^2} \,  \frac{(1+2\cosh\rho)\sinh\rho}{6} \Big( 1+ \frac{\calS}{r^4} \, \frac{\cosh\rho -1}{1 + 2\cosh\rho} \Big) \\
\hat B_{\alpha\beta} &=  \frac{k(r)}{3r^6} \, |\mu + \chi|^4 \cosh^2\frac{\rho}{2} \; M_{\alpha\beta} \;, \qquad\qquad\qquad M_{\alpha\beta} \equiv  \left( \begin{smallmatrix} & & & \!\!\!\! - \sinh\frac{\rho}{2} \\ & & \!\!\!\! \cosh\frac{\rho}{2} \\ & \!\!\!\! - \cosh\frac{\rho}{2} \\ \sinh\frac{\rho}{2} \end{smallmatrix} \right) \;.
\eea
Some useful relations are:
\bea
\sqrt{-\tilde q} &= \sqrt{-\gamma} + g_s e^{-\Phi} h^{-1} \sqrt{\hat B} \qquad\qquad &
\sqrt{-\tilde q} \, \tilde q^{(\alpha\beta)} &= \sqrt{-\gamma} \, \gamma^{\alpha\beta} \\
\tilde q^{\mu\nu} &= h^{1/2}\, \eta^{\mu\nu} = \gamma^{\mu\nu} &
\sqrt{-\tilde q} \, \tilde q^{[\alpha\beta]} &= {g_s}^\frac{1}{2} e^{-\frac{\Phi}{2}} \, h^{-1} \sqrt{\hat B} \, (\hat B^{-1})^{\alpha\beta} \;. \nonumber
\eea
These are valid exactly in the SUSY background, and at leading order in $\cS$ in the DKM solution.

\paragraph{The scalar potential.} $\cL_C$ contains the potential for the scalars. In the SUSY background $\cL_C = 0$, and since it is the only source of possible linear terms, the equations of motion are solved for any value $\chi = const$. In the DKM solution some linear terms survive:
\be
\cL_C \quad\sim\quad \frac{g_s \calS }{|\mu|^2} \, e^{-\rho} \, (\bar\mu \chi + \mu \bar\chi) \;.
\ee
This means that $\chi = \bar\chi = 0$ is no longer a solution. By inspection of the action (\ref{reduced action}), one finds that the equations of motion can be solved by $\phi^i(\rho)$ and $A_{h_2}(\rho)$ and all the other fields consistently set to zero.

\section{Vector mesons spectrum}
\label{app: vector mesons}

With the effective action at hand, we can in principle compute the full KK spectrum on the D7-brane: normalizable modes correspond to 4d mesons of the strongly coupled theory, of the form (\ref{messenger operators}). We will focus on the tower of vector mesons, expecting the results to be qualitatively the same as for the other kind of mesons. This allows us not to solve for the deformed profile $\big( \phi^i(\rho),\, A_{h_2}(\rho) \big)$ in the DKM solution. The result is
\be \label{meson spectrum}
M_n^2 = \frac{|\mu|^{4/3}}{L_\mathrm{eff}^4} \, \lambda_n^2 - \frac{\calS \, \alpha'^4}{|\mu|^{4/3} L_\mathrm{eff}^4} \, \delta \beta_n \;,
\ee
where $\lambda_n$ and $\delta\beta_n$ are positive numbers of order one. Both $\lambda_n$ and $\delta\beta_n/\lambda_n$ turn out to be almost exactly linear functions of $n$.

The ansatz for vector mesons is $A_\mu (x,\rho) = v_\mu e^{ik \cdot x} a(\rho)$, with $-k^2 = M_n^2$ the 4d mass. One can see that the only piece of action relevant for the EOM's of $A_\mu$ is
\be
\label{action vector mesons}
S_\text{vector mesons} = \frac{1}{2\pi^5 g_s} \int d^4x\, d\rho\, e^{\Phi} \, \sqrt{-\tilde q} \, \tilde q^{(ab)} \tilde q^{(cd)} F_{ac} F_{bd} \;.
\ee
Moreover the EOM's of $\big( \phi^i(\rho), A_{h_2}(\rho) \big)$ decouple from those of $A_\mu(x,\rho)$. The EOM's are $\partial_c \big( e^\Phi \sqrt{-\tilde q}\, \tilde q^{(cd)} \tilde q^{(ab)} F_{db} \big) = 0 $. The equation with $a = \rho$ gives $k \cdot v = 0$. The one with $a = \nu$ gives the wavefunction equation
\be \label{diff equation vector mesons}
\big( e^\Phi \sqrt{-\gamma} \, \gamma^{\rho\rho} h^{1/2} a'(\rho) \big)' - k^2 \, e^\Phi \sqrt{-\tilde q} \, h \, a(\rho) = 0 \;,
\ee
where derivatives are taken with respect to $\rho$.

We solve the equation perturbatively in $\cS$: we numerically compute the spectrum and the wavefunctions for $\cS = 0$ (KT solution), and afterwards the mass perturbation induced by $\cS$ (DKM solution). It will be convenient to recast the equation in the form $D a(\rho) = \beta_n \, a(\rho)$, where $D \equiv f_2^{-1} \partial( f_1 \partial)$ and
{\small
\bea
f_1 &= \frac{3\sinh\rho}{(1+\cosh\rho)^{1/3}} \Big[ 1 + \frac{\calS}{r^4} \Big( \frac{\cosh\rho-1}{1 + 2\cosh\rho} - 3 \log \frac{r}{r_0} \Big) \Big] \\
f_2 &= \frac{\sinh\rho}{(1+\cosh\rho)^2} \frac{\log r/r_0}{\log r_\mu/r_0} \Big[ \frac{1+2\cosh\rho}{6} \Big( 1 + \frac{\calS}{2r^4} \, \frac{4\cosh\rho -1}{1 + 2\cosh\rho} - \frac{3\calS}{r^4} \log \frac{r}{r_0} \Big) + \log \frac{r}{r_0} \Big( 1 + \frac{3\calS}{r^4} \Big) \Big] \\
\beta_n &= -\lambda_n^2 = - M_n^2 \, \frac{L_\mathrm{eff}^4}{|\mu|^{4/3}} \;.
\eea
}
Here $r_\mu^3 = 2|\mu|^2$ is the radial location of the D7-brane tip, $L_\mathrm{eff}$ is the effective AdS$_5$ radius at $r_\mu$
\be
L_\mathrm{eff}^4 \equiv \frac{27\pi\alpha'^2}{4}\, \frac{3(g_sM)^2}{2\pi} \, \log\frac{r_\mu}{r_0} \;,
\ee
$r_0$ has been tuned such that $N=0$, and we adopt the approximation $1/4 \ll \log r_\mu/r_0$.

For $\cS = 0$, the regular normalizable solutions of (\ref{diff equation vector mesons}) behave as $a \sim const$ in the IR and $a \sim e^{-2\rho/3}$ in the UV; the spectrum $\lambda_n$ is then obtained with the ``shooting technique''. The result depends on $\log r_\mu/r_0$, and we give some sample numbers in (\ref{KK numbers}).
It turns out that the eigenvalues $\lambda_n$ lay almost exactly along a line: $\lambda_n = c_1 + c_2 \, n$, for coefficients $c_{1,2}$.

All wavefunctions are localized at the tip of the D7-brane, with support of order one in the dimensionless coordinate $\rho$ that grows only logarithmically with $n$. The $n$-th wavefunction with 4d mass $M_n$ has $n-1$ nodes.

For small $\cS$, we evaluate the perturbation operator $D^{(1)}$ on the normalized unperturbed wavefunctions $a^{(0)}$. The result is in (\ref{meson spectrum}), where the parametric dependence is isolated with $\lambda_n$ and $\delta\beta_n$ being numbers of order one. For some particular choice of the mass $|\mu|$ we got:
\bea \label{KK numbers}
&\log r_\mu/r_0 = 1 \;: & &
\begin{array}{c@{\quad=\quad}cccccc}
\lambda_n & 1.83, & 3.73, & 5.55, & 7.37, & 9.18 & \dots \\
\delta\beta_n & 1.00, & 6.72, & 15.3, & 27.5, & 42.9, & \dots
\end{array} \\
&\log r_\mu/r_0 = 100 \;: & &
\begin{array}{c@{\quad=\quad}cccccc}
\lambda_n & 0.293, & 0.724, & 1.15, & 1.58, & 2.01 & \dots \\
\delta\beta & 2.58, & 33.2, & 87.4, & 166, & 268, & \dots
\end{array}
\eea
It turns out that the numbers $\delta\beta_n/\lambda_n$ lay almost exactly along a line. They are proportional to the gaugino mass induced at 1-loop by the $n$-th messenger meson.

An extra worldvolume flux $F_2$ can be taken into account easily. Notice that the expression in (\ref{KT worldvolume flux}) is exactly proportional to $\hat B_2$, therefore the inclusion of $P$ units of flux wrapping $S^2$ amounts to the substitution
\be
k(r) \;\to\; k(r) + \frac{3}{2}\, P \;.
\ee
The eigenmode equation (\ref{diff equation vector mesons}) can then be re-evaluated in the same way.

\section{Supersymmetric vector mesons on the deformed conifold}
\label{app: vector mesons deformed}

As a by-product of our analysis, with slight modifications we can get the spectrum of vector mesons on the massive Kuperstein embedding $z_4 = \mu$ in the KS solution $\sum_{i=1}^4 z_i^2 = \epsilon^2$. First of all we parametrize the holomorphic coordinates as
\bea \label{holomorphic coordinates wp def}
z_1 &= i \, \eta(\chi) \Big[ \cos\phi\, \cosh \Big( \frac{\rho + i\gamma}{2} \Big) \cos\theta - i \sin\phi\, \sinh \Big( \frac{\rho + i\gamma}{2} \Big) \Big] \\
z_2 &= i \, \eta(\chi) \Big[ \sin\phi\, \cosh \Big( \frac{\rho + i\gamma}{2} \Big) \cos\theta + i \cos\phi\, \sinh \Big( \frac{\rho + i\gamma}{2} \Big) \Big] \\
z_3 &= i \, \eta(\chi) \cosh \Big( \frac{\rho + i\gamma}{2} \Big) \sin\theta \\
z_4 &= \mu + \chi \;,
\eea
which describes the deformed conifold as a foliation of Kuperstein embeddings. Here
\be
\eta(\chi) = \sqrt{ (\mu + \chi)^2 - \epsilon^2 } \;.
\ee
It will be useful to define different radial coordinates $r$, $\tau$ and $\rho$:
\be \label{KS radial functions}
\sum_{i=1}^4 |z_i|^2 = r^3 = |\epsilon|^2 \cosh\tau = |\eta|^2 \cosh\rho + |\mu + \chi|^2 \;.
\ee
Then we will use the same left-invariant forms as in (\ref{left-invariant forms}).

The K\"ahler form, warp factor and B-field of the KS solution can be written as
\bea \label{KS solution}
J &= \frac{i K(\tau)}{3|\epsilon|^{2/3}} \Big[ \delta_{ij} - \frac{1}{|\epsilon|^2 \sinh^2\tau} \Big( \cosh\tau - \frac{2}{3K(\tau)^3} \Big) \, \bar z_i \, z_j \Big] \, dz_i \wedge d\bar z_j \\
h &= \frac{9(g_s M \alpha')^2}{2^{4/3} |\epsilon|^{8/3}} \, I(\tau) \qquad\qquad I(\tau) = \int_\tau^\infty dx\,  \frac{x \coth x -1}{\sinh^2 x} \, \big( \sinh 2x \, - 2x\big)^{1/3} \\
B_2 &= b(\tau) \, i\, \epsilon_{ijkl} \, z_i \bar z_j \, dz_k \wedge d\bar z_l \qquad\qquad\qquad b(\tau) = \frac{g_s M \alpha'}{2|\epsilon|^4} \, \frac{\tau \coth\tau -1}{\sinh^2\tau} \;,
\eea
where
\be
K(\tau)^3 = \frac{\sinh 2\tau - 2\tau}{2 \sinh^3 \tau} \;.
\ee
The unwarped metric is easily derived as $ds_6^2 = -2i J_{i \bar \jmath} dz_i \otimes d\bar z_j$. For our computation of the vector meson spectrum, we will only need the unwarped pulled-back metric and the B-field determinant:
\bea
ds_4^2 \Big|_{\chi = 0} &= \frac{K(\tau) \, |\eta|^2}{2 |\epsilon|^{2/3}} \, \Big[ K_2(\rho) \big( d\rho^2 + h_3^2 \big) + \cosh^2\frac{\rho}{2}\, h_1^2 + \sinh^2\frac{\rho}{2}\, h_2^2 \Big] \\
\sqrt{\det \hat B} &= b(\tau)^2 |\eta|^4 \frac{\sinh\rho}{2} \Big[ \cosh^4\frac{\rho}{2} \, |\eta\bar\mu|^2 - \big( \im \, \eta\bar\mu \big)^2 \cosh\rho \Big] \;,
\eea
where
\be
K_2(\rho) = \cosh\rho - \frac{|\eta|^2 \sinh^2\rho}{|\epsilon|^2 \sinh^2\tau} \Big( \cosh\tau - \frac{2}{3K(\tau)^3} \Big) \;,
\ee
$\eta = \sqrt{\epsilon^2 - \mu^2}$ is simply a parameter and remember that $\tau = \tau(\rho)$ according to (\ref{KS radial functions}).

As in section \ref{app: vector mesons}, the eigenmode equation for vector mesons is
\be
\big( \sqrt{-\gamma} \, \gamma^{\rho\rho} h^{1/2} a'(\rho) \big)' - k^2 \, \sqrt{-\tilde q} \, h \, a(\rho) = 0 \;,
\ee
and one has to impose regularity at the origin $a'(0)=0$ and normalizability at infinity $a(\infty) = 0$. Given the pieces
\be \nonumber
\sqrt{-\gamma} = \frac{|\eta|^4 K(\tau)^2 K_2(\tau) \sinh\rho}{72 \, |\epsilon|^{4/3}} \qquad\qquad\qquad
\sqrt{-\gamma} \, \gamma^{\rho\rho} \, h^{1/2} = \frac{|\eta|^2 K(\tau) \sinh\rho}{12 \, |\epsilon|^{2/3}} \;,
\ee
the equation can then be solved numerically.

Let us just mention two obvious limit cases. When $\epsilon \to 0$, we must take $\tau \to \infty$ with $|\epsilon|^2 e^\tau = 2 r^3$ and use the radial coordinate $r$ instead of $\tau$. Then $\eta = \mu$,
\be \nonumber
K(\tau) \;\to\; |\epsilon|^{2/3}/r \:, \qquad\quad K_2(\tau) \;\to\; (1 + 2\cosh\rho)/3 \:, \qquad\quad b(\tau) \;\to\; k(r)/3r^6 \;,
\ee
and the warp factor goes into the KT one. We exactly reproduce the supersymmetric result of section \ref{app: vector mesons}, namely a KK spacing of order $M_1 \sim |\mu|^{2/3}/L_\mathrm{eff}^2$. On the other hand, when $\mu = 0$ we get $\eta = i\epsilon$ and $\tau = \rho$. Moreover $K_2(\tau) = 2/(3K^3(\tau))$ and $\hat B_2 = 0$. The equation simplifies to
\be
\big[ (\sinh 2\tau - 2\tau)^{1/3} a' \big]' - k^2 \frac{(g_s M \alpha')^2}{2^{2/3}|\epsilon|^{4/3}} \, \frac{I(\tau) \sinh^2\tau}{(\sinh 2\tau - 2\tau)^{1/3}} \, a = 0 \;.
\ee
In this case the KK spacing is of order $M_1 \sim |\epsilon|^{2/3}/L^2 \sim m_\mathrm{glueball}$. A similar computation has been performed in \cite{Kuperstein}.

\section{Slowing down the running}
\label{app: slowing down}

It could be useful to introduce some additional parameter to slow down the RG flow and extend the throat, relaxing the bound on the number $k$ of cascade steps derived from (\ref{running_alpha}). One possible way is by orbifolding. From a gravity perspective this is a natural choice, since it reduces the volume of the D7-branes.

\begin{figure}[tn]
\begin{center}
\psfrag{LIR}[cc][][1]{$\epsilon$}
\psfrag{mu}[cc][][1]{$\mu$}
\psfrag{a}[cc][][1]{$\alpha^{-1}$}
\psfrag{log(L)}[cc][][1]{$\log \Lambda$}
\includegraphics[width=9cm]{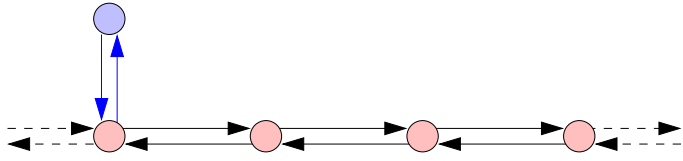}
\caption{Quiver diagram for the $\mathbb{Z}_2$ orbifold of the conifold with flavors. The quiver has 4 nodes and is periodically identified.}
\label{orbifold_conifold}
\end{center}
\end{figure}

For concreteness, let us discuss the $\bbZ_Q$ non-chiral orbifold of the conifold:
\be
z_1 + i z_2 \;\to\; e^{2\pi i/Q} \, (z_1 + i z_2) \;, \qquad z_1 - i z_2 \;\to\; e^{-2\pi i/Q} \, (z_1 - i z_2) \;,
\ee
which is described by the equation $xy = (zw)^Q$ \cite{Uranga:1998vf}. Suppose to start with the KT background with $N$ units of 5-form flux at some radius $r_\mr{down}$ and $M$ units of 3-form flux (with both $N$ and $M$ multiples of $Q$), and with $K$ D7-branes along the Kuperstein embedding $z_4 = \mu$. The dual field theory group is $SU(N+M) \times SU(N)$, with the weakly gauged MSSM $SU(K)$ group attached to, say, $SU(N+M)$. After the orbifold projection we have $N/Q$ D3 and $M/Q$ D5-branes on the physical space (as opposed to $N$ and $M$ on the covering space), and still $K$ D7-branes along $z+w = 2i\mu$. The dual field theory has $2Q$ nodes, half of which representing $SU\big( (N+M)/Q\big)$ and half $SU(N/Q)$, with the MSSM group attached to one $SU\big( (N+M)/Q \big)$; see figure \ref{orbifold_conifold}. Notice that the warp factor as well as the AdS$_5$ radius and the curvature are the same as in the original KT background.

As in section \ref{sec: geometric constraints}, we define ``period'' a piece of RG flow along which all nodes undergo one Seiberg duality. Its radial length is
\be
r_\mr{up} = r_\mr{down} \, e^{4\pi \over 3g_s M} \;,
\ee
the same as in the original KT background. At scale $r_\mr{up}$ the theory has $Q$ $SU\big((N+3M)/Q \big)$ nodes and $Q$ $SU\big( (N+2M)/Q \big)$ nodes. During the whole piece of flow, the messengers contribute $\frac{N+M}{Q}$ flavors to the MSSM group and then its $\beta$-function is
\be
\beta = - \frac{3}{2} \Big( 2 \Big[ \frac{k}{2} \Big]_- + 1 \Big) \frac{M}{Q} \;,
\ee
where we introduced the notation $N = kM$ and $k$ counts the would-be number of Seiberg dualities in the {\it original} KT background -- it is related to the radial coordinate by $k / k_\mr{IR} = (3g_sM/2\pi) \log \big( r/r_{IR} \big)$, as in (\ref{Lambda_k}). Therefore the running of the gauge coupling during one period is
\be
\Delta x \equiv \Delta \Big( \frac{8\pi^2}{g^2} \Big) = \frac{2\pi \big( 2[k/2]_- + 1 \big)}{g_s Q} \;.
\ee
After summing over the periods ($k$ increases by 2 from period to period) we get the same formula as in (\ref{running_alpha}), but reduced by a factor of $Q$:
\be
\alpha^{-1}_{\rm IR} - \alpha^{-1}_{\rm UV} \sim \frac{1}{4 g_s Q} (k_{\rm UV}^2 - k_{\rm IR}^2) \;.
\ee
This result is easily rederived in supergravity: when reading the MSSM gauge coupling from the DBI action, the induced metric and warp factor are the same as in the original KT background, but we only integrate over a $1/Q$ fraction of the coordinate range.

If we apply the orbifold projection to the DKM solution, we find that the vacuum energy is reduced by a factor of $Q$ (while the scale $\epsilon$ and the hierarchy are not modified), because the number of anti-D3-branes on the physical space is $p/Q$. Moreover the vector meson spectrum is not modified, because in our DBI plus WZ computation both the kinetic and mass terms are reduced by a factor of $Q$ (for instance the whole action in (\ref{action vector mesons})). Therefore, even though a more refined analysis is desirable, orbifolding appears to be a promising way to avoid possibly annoying constraints.

\newpage

\end{document}

%% file: texpref.tex
\newcommand{\eref}[1]{(\ref{#1})}

\newcommand{\fref}[1]{Figure~\ref{#1}}
\newcommand{\cref}[1]{Chapter~\ref{#1}}
\newcommand{\beq}{\begin{equation}}
\newcommand{\eeq}{\end{equation}}
\newcommand{\ba}{\begin{array}}
\newcommand{\ea}{\end{array}}
\newcommand{\bcenter}{\begin{center}}
\newcommand{\ecenter}{\end{center}}

\def\IB{\relax\hbox{$\inbar\kern-.3em{\rm B}$}}
\def\IC{\relax\hbox{$\inbar\kern-.3em{\rm C}$}}
\def\ID{\relax\hbox{$\inbar\kern-.3em{\rm D}$}}
\def\IE{\relax\hbox{$\inbar\kern-.3em{\rm E}$}}
\def\IF{\relax\hbox{$\inbar\kern-.3em{\rm F}$}}
\def\IG{\relax\hbox{$\inbar\kern-.3em{\rm G}$}}
\def\IGa{\relax\hbox{${\rm I}\kern-.18em\Gamma$}}
\def\IH{\relax{\rm I\kern-.18em H}}
\def\IK{\relax{\rm I\kern-.18em K}}
\def\IL{\relax{\rm I\kern-.18em L}}
\def\IP{\relax{\rm I\kern-.18em P}}
\def\IR{\relax{\rm I\kern-.18em R}}
\def\IZ{\relax\ifmmode\mathchoice
{\hbox{\cmss Z\kern-.4em Z}}{\hbox{\cmss Z\kern-.4em Z}}
{\lower.9pt\hbox{\cmsss Z\kern-.4em Z}}
{\lower1.2pt\hbox{\cmsss Z\kern-.4em Z}}\else{\cmss Z\kern-.4em Z}\fi}
\def\II{\relax{\rm I\kern-.18em I}}

\def\sCC{{\kern 0.27em\vrule height1.45ex width0.03em depth0em
          \kern-0.30em\rm C}}
\def\C{{\mathchoice
  {\sCC}
  {\sCC}
  {\kern 0.225em \vrule height1.05ex width0.025em depth0em \kern-0.25em \rm C}
  {\kern 0.180em \vrule height0.78ex width0.02em depth0em \kern-0.2em \rm C}
        }}
\def\sHH{{\rm I\kern-.16em{}H}}
\def\H{{\mathchoice
  {\sHH}
  {\sHH}
  {\rm I\kern-.13em{}H}
  {\rm I\kern-.13em{}H} }}
\def\sNN{{\rm I\kern-.16em{}N}}
\def\N{{\mathchoice
  {\sNN}
  {\sNN}
  {\rm I\kern-.12em{}N}
  {\rm I\kern-.10em{}N} }}
\def\sPP{{\rm I\kern-.16em{}P}}
\def\P{{\mathchoice
  {\sPP}
  {\sPP}
  {\rm I\kern-.12em{}P}
  {\rm I\kern-.10em{}P} }}
\def\sQQ{{\kern 0.27em \vrule height1.45ex width0.03em depth0em
          \kern-0.30em \rm Q}}
\def\Q{{\mathchoice
        {\sQQ}
        {\sQQ}
  {\kern 0.225em \vrule height1.05ex width0.025em depth0em \kern-0.25em \rm Q}
  {\kern 0.180em \vrule height0.78ex width0.020em depth0em \kern-0.20em \rm Q}
        }}
\def\sRR{{\rm I\kern-0.16em{}R}}
\def\R{{\mathchoice
  {\sRR}
  {\sRR}
  {\rm I\kern-0.12em{}R}
  {\rm I\kern-0.10em{}R} }}
\def\sZZ{{\rm Z\kern-0.32em{}Z}}
\def\Z{{\mathchoice
  {\sZZ}
  {\sZZ} 
  {\rm Z\kern-0.3em{}Z}     
  {\rm Z\kern-0.25em{}Z} }}  
\def\ZZZ{{\rm Z\kern-0.24em{}Z}}
\def\sII{{\rm I\kern-0.16em{}I}}
\def\I{{\mathchoice
  {\sII}
  {\sII}
  {\rm I\kern-0.12em{}I}
  {\rm I\kern-0.10em{}I} }}

\def\Tr{{\rm Tr}}

\def\inbar{\,\vrule height1.5ex width.4pt depth0pt}
\font\cmss=cmss10 \font\cmsss=cmss10 at 7pt

\def\smiley{\hbox{\large$\bigcirc$\hspace{-0.80em}\raise.2ex
\hbox{$\cdot\cdot$}\kern-.61em\lower.2ex\hbox{\scriptsize$\smile$}}\ }
\def\frowny{\hbox{\large$\bigcirc$\hspace{-0.80em}\raise.2ex
\hbox{$\cdot\cdot$}\kern-.635em\lower.2ex\hbox{\scriptsize$\frown$}}\ }

\def\I{{\rlap{1} \hskip 1.6pt \hbox{1}}}

\makeatletter
\let\hangafter\@hangfrom
\makeatother

%% file: paper.bbl
\begin{thebibliography}{10}

\bibitem{classic}
M. Dine, W. Fischler and M. Srednicki, ``Supersymmetric techicolor," Nucl. Phys. {\bf B189} (1981) 575;
S. Dimopoulos and S. Raby, ``Supercolor," Nucl. Phys. {\bf B192} (1981) 353;
M. Dine and W. Fischler, ``A phenomenological model of particle physics based on supersymmetry,"
Phys. Lett. {\bf B110} (1982) 227; C. Nappi and B. Ovrut, ``Supersymmetric extension of the
$SU(3) \times SU(2) \times U(1)$ model,"  Phys. Lett. {\bf B113} (1982) 175;
L. Alvarez-Gaume, M. Claudson and M. Wise, ``Low-energy supersymmetry,"
Nucl. Phys. {\bf B207} (1982) 96; S. Dimopoulos and S. Raby, ``Geometric hierarchy,"
Nucl. Phys. {\bf B219} (1983) 479.

\bibitem{classictwo}
M. Dine, A. Nelson and Y. Shirman, ``Low-energy dynamical supersymmetry breaking
simplified," Phys. Rev. {\bf D51} (1995) 1362 [arXiv:hep-ph/9408384];
M. Dine, A. Nelson, Y. Nir and Y. Shirman, ``New tools for low-energy dynamical
supersymmetry breaking," Phys. Rev. {\bf D53} (1996) 2658 [arXiv:hep-ph/9507378].

\bibitem{GRreview}
G.F. Giudice and R. Rattazzi, ``Theories with gauge mediated supersymmetry breaking,"
Phys. Rept. {\bf 322} (1999) 419 [arXiv:hep-ph/9802171].

\bibitem{GGD}
O. Aharony, S. Gubser, J. Maldacena, H. Ooguri and Y. Oz, ``Large N field theories,
string theory and gravity," Phys. Rept. {\bf 323} (2000) 183 [arXiv:hep-th/9905111].


\bibitem{meade-et-al}
P. Meade, N. Seiberg and D. Shih, ``General Gauge Mediation," arXiv:0801.3278 [hep-ph].

\bibitem{Verlinde:2007qk}
  H.~Verlinde, L.~T.~Wang, M.~Wijnholt and I.~Yavin,
  ``A Higher Form (of) Mediation,''
  JHEP {\bf 0802}, 082 (2008)
  [arXiv:0711.3214 [hep-th]].

\bibitem{Dermisek:2007qi}
  R.~Dermisek, H.~Verlinde and L.~T.~Wang,
  ``Hypercharged Anomaly Mediation,''
  Phys.\ Rev.\ Lett.\  {\bf 100}, 131804 (2008)
  [arXiv:0711.3211 [hep-ph]].

\bibitem{Grimm:2008ed}
  T.~W.~Grimm and A.~Klemm,
  ``U(1) Mediation of Flux Supersymmetry Breaking,''
  JHEP {\bf 0810}, 077 (2008)
  [arXiv:0805.3361 [hep-th]].

\bibitem{KS}
I.R. Klebanov and M.J. Strassler, ``Supergravity and a confining gauge theory: Duality cascades and $\chi$SB
resolution of naked singularities," JHEP {\bf 0008} (2000) 052 [arXiv:hep-th/0007191].

\bibitem{Kachru:2002gs}
  S.~Kachru, J.~Pearson and H.~L.~Verlinde,
    ``Brane/flux annihilation and the string dual of a non-supersymmetric  field
      theory,''
        JHEP {\bf 0206}, 021 (2002)
	  [arXiv:hep-th/0112197].

\bibitem{semi-direct}
N. Seiberg, T. Volansky and B. Wecht, ``Semi-direct Gauge Mediation," JHEP {\bf 0811} (2008) 004 [arXiv:0809.4437].

\bibitem{gaugino}
D.E. Kaplan, G. Kribs and M. Schmaltz, ``Supersymmetry breaking through transparent extra dimensions," Phys. Rev.
{\bf D62} (2000) 035010 [arXiv:hep-ph/9911293]; Z. Chacko, M. Luty, A.E. Nelson and E. Ponton, ``Gaugino mediated
supersymmetry breaking," JHEP {\bf 0001} (2000) 003 [arXiv:hep-ph/9911323].

\bibitem{Nomura}
W. Goldberger, Y. Nomura and D. Tucker-Smith, ``Warped supersymmetric grand unification," Phys. Rev. {\bf D67} (2003) 075021
[arXiv:hep-ph/0209158];
Y. Nomura and D. Tucker-Smith, ``Spectrum of TeV particles in warped supersymmetric grand unification," Phys. Rev. {\bf D68} (2003)
075003 [arXiv:hep-ph/0305214];
Y. Nomura, ``Supersymmetric unification in warped space," arXiv:hep-ph/0410348.

\bibitem{toappear}
F. Benini, A. Dymarsky, S. Franco, S. Kachru, D. Simic and H. Verlinde, to appear.


\bibitem{singlesector}
N.~Arkani-Hamed, M.~A.~Luty and J.~Terning,
  ``Composite quarks and leptons from dynamical supersymmetry breaking  without
    messengers,''
      Phys.\ Rev.\  D {\bf 58}, 015004 (1998)
        [arXiv:hep-ph/9712389];
  M.~A.~Luty and J.~Terning,
    ``Improved single sector supersymmetry breaking,''
      Phys.\ Rev.\  D {\bf 62}, 075006 (2000)
        [arXiv:hep-ph/9812290].

\bibitem{GGG}
M.~Gabella, T.~Gherghetta and J.~Giedt,
  ``A gravity dual and LHC study of single-sector supersymmetry breaking,''
    Phys.\ Rev.\  D {\bf 76}, 055001 (2007)
      [arXiv:0704.3571 [hep-ph]].


\bibitem{Lippert}
 A.~R.~Frey, M.~Lippert and B.~Williams,
  ``The fall of stringy de Sitter,''
  Phys.\ Rev.\  D {\bf 68}, 046008 (2003)
  [arXiv:hep-th/0305018].

\bibitem{Freivogel}
B. Freivogel and M. Lippert, ``Evidence for a bound on the lifetime of de Sitter space," arXiv:0807.1104 [hep-th].

\bibitem{Oliver}
C.M. Brown and O. DeWolfe, ``Brane/flux annihilation transitions and nonperturbative
moduli stabilization,"  arXiv:0901.4401 [hep-th].


\bibitem{DKM}
  O.~DeWolfe, S.~Kachru and M.~Mulligan,
  ``A Gravity Dual of Metastable Dynamical Supersymmetry Breaking,''
  Phys.\ Rev.\  D {\bf 77}, 065011 (2008)
  [arXiv:0801.1520 [hep-th]].



\bibitem{Ouyang}
P. Ouyang, ``Holomorphic D7 branes and flavored ${\cal N}=1$ gauge theories," Nucl. Phys. {\bf B699} (2004) 207
[arXiv:hep-th/0311084].

\bibitem{Kuperstein}
S. Kuperstein, ``Meson spectroscopy from holomorphic probes on the warped deformed conifold," JHEP {\bf 0503} (2005) 014.

\bibitem{OWitten}
I.~R.~Klebanov, P.~Ouyang and E.~Witten,
  ``A gravity dual of the chiral anomaly,''
  Phys.\ Rev.\  D {\bf 65}, 105007 (2002)
  [arXiv:hep-th/0202056].


\bibitem{KT}
  I.R. Klebanov and A. Tseytlin,
  ``Gravity duals of supersymmetric $SU(N) \times SU(N+M)$ gauge theories,"
  Nucl. Phys. {\bf B578} (2000) 123
  [arXiv:hep-th/0002159].

\bibitem{LO}
  T. Levi and P. Ouyang,
  ``Mesons and flavor on the conifold,"
  Phys. Rev. {\bf D76} (2007) 105022
  [arXiv:hep-th/0506021].

\bibitem{Benini}
  F.~Benini, F.~Canoura, S.~Cremonesi, C.~Nunez and A.~V.~Ramallo,
  ``Backreacting Flavors in the Klebanov-Strassler Background,''
  JHEP {\bf 0709}, 109 (2007)
  [arXiv:0706.1238 [hep-th]].

\bibitem{Benini:2007kg}
  F.~Benini,
  ``A chiral cascade via backreacting D7-branes with flux,''
  JHEP {\bf 0810}, 051 (2008)
  [arXiv:0710.0374 [hep-th]].

\bibitem{Franco}
  S.~Franco, D.~Rodriguez-Gomez and H.~Verlinde,
  ``N-ification of Forces: A Holographic Perspective on D-brane Model
  Building,''
  arXiv:0804.1125 [hep-th].

\bibitem{Cascales:2005rj}
  J.~F.~G.~Cascales, F.~Saad and A.~M.~Uranga,
  ``Holographic dual of the standard model on the throat,''
  JHEP {\bf 0511}, 047 (2005)
  [arXiv:hep-th/0503079].

\bibitem{Argurio:2008mt}
  R.~Argurio, F.~Benini, M.~Bertolini, C.~Closset and S.~Cremonesi,
  ``Gauge/gravity duality and the interplay of various fractional branes,''
  Phys.\ Rev.\  D {\bf 78}, 046008 (2008)
  [arXiv:0804.4470 [hep-th]].

\bibitem{KMS}
  S.~Kachru, L.~McAllister and R.~Sundrum,
  ``Sequestering in string theory,''
  JHEP {\bf 0710}, 013 (2007)
  [arXiv:hep-th/0703105].

\bibitem{Polch_Peet}
 A.~Peet and J.~Polchinski,
   ``UV/IR Relations in AdS Dynamics,''
     Phys.\ Rev.\  D {\bf 59}, 065011 (1999)
       [arXiv:hep-th/9809022].

\bibitem{Sussk_Witten}
  L.~Susskind and E.~Witten,
  ``The Holographic Bound in anti-de Sitter space,''
  Phys.\ Rev.\  D {\bf 59}, 065011 (1999)
  [arXiv:hep-th/9805114].

\bibitem{Kuperstein:2003yt}
  S.~Kuperstein and J.~Sonnenschein,
  ``Analytic non-supersymmetric background dual of a confining gauge theory
  and the corresponding plane wave theory of hadrons,''
  JHEP {\bf 0402}, 015 (2004)
  [arXiv:hep-th/0309011].

\bibitem{CIU}
  P.~G.~Camara, L.~E.~Ibanez and A.~M.~Uranga,
  ``Flux-induced SUSY-breaking soft terms,''
  Nucl.\ Phys.\  B {\bf 689}, 195 (2004)
  [arXiv:hep-th/0311241].

\bibitem{Weinberg_III}
  S.~Weinberg,
  ``The Quantum Theory of Fields, Volume III: Supersymmetry,''
  Cambridge University Press, 2005. 

\bibitem{Martucci}
  L.~Martucci, J.~Rosseel, D.~Van den Bleeken and A.~Van Proeyen,
  ``Dirac actions for D-branes on backgrounds with fluxes,''
  Class.\ Quant.\ Grav.\  {\bf 22}, 2745 (2005)
  [arXiv:hep-th/0504041].

\bibitem{MMS}
  F.~Marchesano, P.~McGuirk and G.~Shiu,
  ``Open String Wavefunctions in Warped Compactifications,''
  arXiv:0812.2247 [hep-th].

\bibitem{MN}
  J.~M.~Maldacena and H.~S.~Nastase,
  ``The supergravity dual of a theory with dynamical supersymmetry  breaking,''
  JHEP {\bf 0109}, 024 (2001)
  [arXiv:hep-th/0105049].

\bibitem{DKV}
  O.~DeWolfe, S.~Kachru and H.~L.~Verlinde,
  ``The giant inflaton,''
  JHEP {\bf 0405}, 017 (2004)
  [arXiv:hep-th/0403123].

\bibitem{Komargodski:2009jf}
  Z.~Komargodski and D.~Shih,
  ``Notes on SUSY and R-Symmetry Breaking in Wess-Zumino Models,''
  arXiv:0902.0030 [hep-th].

\bibitem{Vafa}
  J. Heckman and C. Vafa,  ``From F-theory GUTs to the LHC,"  arXiv:0809.3452 [hep-ph].

\bibitem{Becker:1995kb}
  K.~Becker, M.~Becker and A.~Strominger,
  ``Five-Branes, Membranes And Nonperturbative String Theory,''
  Nucl.\ Phys.\  B {\bf 456}, 130 (1995)
  [arXiv:hep-th/9507158].

\bibitem{Marino:1999af}
  M.~Marino, R.~Minasian, G.~W.~Moore and A.~Strominger,
  ``Nonlinear instantons from supersymmetric p-branes,''
  JHEP {\bf 0001}, 005 (2000)
  [arXiv:hep-th/9911206].

\bibitem{Gomis:2005wc}
  J.~Gomis, F.~Marchesano and D.~Mateos,
  ``An open string landscape,''
  JHEP {\bf 0511}, 021 (2005)
  [arXiv:hep-th/0506179].

\bibitem{'tHooft:1973jz}
  G.~'t Hooft,
  ``A planar diagram theory for strong interactions,''
  Nucl.\ Phys.\  B {\bf 72}, 461 (1974);
  ``A Two-Dimensional Model For Mesons,''
  Nucl.\ Phys.\  B {\bf 75}, 461 (1974).
  E.~Witten,
  ``Baryons In The 1/N Expansion,''
  Nucl.\ Phys.\  B {\bf 160}, 57 (1979).

\bibitem{Kruczenski:2003be}
  M.~Kruczenski, D.~Mateos, R.~C.~Myers and D.~J.~Winters,
  ``Meson spectroscopy in AdS/CFT with flavour,''
  JHEP {\bf 0307}, 049 (2003)
  [arXiv:hep-th/0304032].

\bibitem{Uranga:1998vf}
  A.~M.~Uranga,
  ``Brane Configurations for Branes at Conifolds,''
  JHEP {\bf 9901}, 022 (1999)
  [arXiv:hep-th/9811004].


\end{thebibliography}
